\documentstyle[emulateapj,onecolfloat,psfig]{article}

\long\def\comment#1{}

\def\la{\hbox{ \raise.35ex\rlap{$<$}\lower.6ex\hbox{$\sim$}\ }}
\def\ga{\hbox{ \raise.35ex\rlap{$>$}\lower.6ex\hbox{$\sim$}\ }}

\def\W2{{\cal W}}

\newcommand{\wj}{\left(
		  	  \begin{array}{ccc}
			  l_1  &  l_2  & l_3 \\
		       	    0  &  0    &  0
                          \end{array}
                          \right)}
\newcommand{\wjm}{\left(
			   \begin{array}{ccc}
	 l_1 & l_2  & l_3  \\
         m_1 & m_2  & m_3
		           \end{array}
                   \right)}
\newcommand{\wjmp}[3]{\left(
			   \begin{array}{ccc}
	 l_#1 & l_#2  & l_#3  \\
         m_#1 & m_#2  & m_#3
		           \end{array}
                   \right)}
\newcommand{\wjma}[6]{\left(
			   \begin{array}{ccc}
	 #1 & #2  & #3  \\
         #4 & #5  & #6
		           \end{array}
                   \right)}
\newcommand{\bi}{B_{l_1 l_2 l_3}}
\newcommand{\bip}{B_{l_1' l_2' l_3'}}
\newcommand{\deld}{\delta^{\rm D}}
\newcommand{\bn}{\hat{\bf n}}
\newcommand{\bm}{\hat{\bf m}}
\newcommand{\bl}{\hat{\bf l}}
\newcommand{\bk}{\hat{\bf k}}
\newcommand{\veck}{{\bf k}}
\newcommand{\rad}{r} 	
\newcommand{\da}{d_A}	
\newcommand{\tableskip}{\tablevspace{3pt}}
\newcommand{\dop}{{\rm dop}}
\newcommand{\sz}{{\rm SZ}}
\newcommand{\isw}{{\rm ISW}}
\newcommand{\sw}{{\rm SW}}
\newcommand{\ov}{{\rm OV}}
\newcommand{\se}{{\rm S}}
\newcommand{\ri}{{\rm ri}}

\newcommand{\Ylm}[1]{Y_{l_#1}^{m_#1}}
\newcommand{\Ylmn}{Y_{l}^{m}}
\newcommand{\alm}[1]{a_{l_#1 m_#1}}
\newcommand{\almn}{a_{l m}}

\begin{document}
\twocolumn[
\title{Imprint of Reionization on the Cosmic Microwave
Background Bispectrum}   
\author{Asantha Cooray$^1$ and Wayne Hu$^2$}
\affil{
$^1$Department of Astronomy and Astrophysics, University of Chicago,
Chicago IL 60637\\
$^2$Institute for Advanced Study, Princeton, NJ 08540\\
E-mail: asante@hyde.uchicago.edu, whu@ias.edu}
\begin{abstract}
We study contributions to the cosmic microwave background (CMB) 
bispectrum from non-Gaussianity induced by secondary anisotropies during 
reionization.  Large-scale structure in the reionized epoch both gravitational
lenses CMB photons and produces Doppler shifts in their temperature from 
scattering off electrons in infall. The resulting correlation is potentially
observable through the CMB bispectrum.  The second-order Ostriker-Vishniac also 
couples to a variety of linear secondary effects to produce a
bispectrum.  For the currently favored flat cosmological model with a low matter content and small optical 
depth in the reionized epoch $\tau \la 0.3$, however, these bispectrum contributions 
are well below the detection threshold of MAP and at or below that of Planck,
given their cosmic and noise variance limitations. 
At the upper end of this range, they can serve as an extra source of
noise for measurements with Planck of either primordial nongaussianity or
that induced by the correlation of gravitational lensing with
the integrated Sachs-Wolfe and the thermal Sunyaev-Zel'dovich
effects.  We include a discussion of the general properties 
of the CMB bispectrum,
its configuration dependence for the various effects,
and its computation in the Limber approximation and beyond.
\end{abstract}


]

\section{Introduction}

The increase in sensitivity of upcoming cosmic microwave background
(CMB) experiments, especially
satellite missions, raise the possibility that
higher order correlations
 in the CMB temperature fluctuations, beyond the two-point function, may be
experimentally detected and studied in detail to look for
deviations from Gaussianity. In the absence deviations 
in the initial conditions, they may arise from the imprint of the
non-linear growth of structures on secondary anisotropies.

Early theoretical work on the three-point
correlation function was aimed at distinguishing between various
theories of the origin of the fluctuations 
(e.g., \cite{Falk93} 1993; \cite{LuoSch93} 1993; 
\cite{Luo94} 1994; \cite{Ganetal94} 1994).
Within the context of inflationary models, especially within 
slow-roll inflation, this topic has been
further addressed in several recent papers (e.g., \cite{GanMar99}
1999; \cite{WanKam99} 1999).
In typical models, the expected non-Gaussian contribution
is below the level that can be detected due to cosmic variance
limitations.

Recently, there has been much renewed interest in the three point
function and its Fourier analogue the bispectrum,
both in anticipation of the high precision satellite data and
from analyses of the COBE DMR data
(e.g., \cite{Hinetal95} 1995; \cite{Feretal98} 1998; \cite{Pando98} 1998).
The detection reported by \cite{Feretal98} (1998) and \cite{Pando98} (1998) has been subsequently
shown to be associated with a known systematic error in the
data time-stream (\cite{Banetal99} 1999; see also \cite{BroTeg99} 1999).
This should be taken as a cautionary tale that even if an
experiment has the raw sensitivity to detect expected bispectrum
signals, removing systematic effects to the required level will prove
challenging indeed.

Nonetheless, the upcoming CMB experiments, especially
MAP\footnote{http://map.nasa.gsfc.gov} and Planck
surveyor\footnote{http://astro.estec.esa.nl/Planck/; also, ESA 
D/SCI(6)3.}, with their higher sensitivity and angular
resolution can in principle be used to study the bispectrum.
It is therefore important to investigate what signals
are expected in the currently favored cosmological models and how
they might be separated from each other.
Recently \cite{SpeGol99}
(1999) and \cite{GolSpe99} (1999) studied the CMB bispectrum under
the context of non-Gaussian contribution due second order gravitational
effects, notably the correlation induced by gravitational lensing of
CMB photons and secondary anisotropies from  the integrated 
Sachs-Wolfe (ISW;
\cite{SacWol67} 1967) effect 
and the thermal Sunyaev-Zel'dovich (SZ; \cite{SunZel80} 1980) effect. 

Here, we present two additional effects resulting through
reionization. The first one is the correlation induced by lensing
through the secondary Doppler effect. 
The second one arises  from the coupling of 
the second order Ostriker-Vishniac effect (OV; Ostriker \&
Vishniac 1986a,b; \cite{Vis87} 1987) to other linear secondary effects 
such as the ISW, Doppler and SZ effects. 
The resulting bispectrum is proportional to the square of the matter
density power spectra. 
We also study the configuration dependence of these effects to address
means by which they may be separated from each other and any primordial
nongaussianity.  

The layout of the paper is as follows.
In \S~\ref{sec:formalism}, we review the background material relevant
for understanding the CMB bispectrum and its statistical properties
in the context of the adiabatic cold dark matter (CDM) models.
A summary of useful properties of the Wigner-3$j$ symbol is presented
in the Appendix.
In \S~\ref{sec:lensing}, we detail the coupling between gravitational
lensing angular excursions and secondary anisotropies. 
Results for the currently favored cosmology are presented in
\S~\ref{sec:lensingresults}.
In \S~\ref{sec:ostrikervishniac}, we treat the
coupling between Ostriker-Vishniac (OV) effect and secondary temperature
fluctuations; results for the favored cosmological model 
are presented in \S \ref{sec:ovresults}. 
We conclude in \S \ref{sec:discussion}.

\section{Preliminaries}
\label{sec:formalism}

We first review the properties of adiabatic CDM models relevant to
the present calculations. We then discuss the general 
properties of the angular bispectrum of the CMB.

\subsection{Adiabatic CDM Model}
The expansion rate for adiabatic CDM cosmological models with a
cosmological constant is
\begin{equation}
H^2 = H_0^2 \left[ \Omega_m(1+z)^3 + \Omega_K (1+z)^2
              +\Omega_\Lambda \right]\,,
\end{equation}
where $H_0$ can be written as the inverse
Hubble distance today $H_0^{-1} = 2997.9h^{-1} $Mpc.
We follow the conventions that 
in units of the critical density $3H_0^2/8\pi G$,
the contribution of each component is denoted $\Omega_i$,
$i=c$ for the CDM, $b$ for the baryons, $\Lambda$ for the cosmological
constant. We also define the 
auxiliary quantities $\Omega_m=\Omega_c+\Omega_b$ and
$\Omega_K=1-\sum_i \Omega_i$, which represent the matter density and
the contribution of spatial curvature to the expansion rate
respectively.

Convenient measures of distance and time include the conformal
distance (or lookback time) from the observer
at redshift $z=0$
\begin{equation}
\rad(z) = \int_0^z {dz' \over H(z')} \,,
\end{equation}
and the analogous angular diameter distance
\begin{equation}
\da = H_0^{-1} \Omega_K^{-1/2} \sinh (H_0 \Omega_K^{1/2} \rad)\,.
\end{equation}
Note that as $\Omega_K \rightarrow 0$, $\da \rightarrow \rad$
and we define $\rad(z=\infty)=\rad_0$.

The adiabatic CDM model possesses a
power spectrum of density fluctuations
\begin{equation}
\left< \delta({\bf k})^*\delta({\bf k')} \right> = (2\pi)^3
	\deld({\bf k}-{\bf k'}) {2 \pi^2 \over k^3} \Delta^2(k)\,,
\end{equation}
where $\deld$ is the Dirac delta function\footnote{We also use this
symbol for Kronecker deltas.} and
\begin{equation}
\Delta^{2}(k) = \delta_H^2 \left({k \over
H_0} \right)^{n+3}T^2(k) \,,
\end{equation}
in linear perturbation theory.  We use
the fitting formulae of \cite{EisHu99} (1999) in evaluating the
transfer function $T(k)$ for CDM models.
Here, $\delta_H$ is the amplitude of present-day density fluctuations
at the Hubble scale; we adopt the COBE normalization for
$\delta_H$ (\cite{BunWhi97} 1997).

The density field may be scaled backwards to higher redshift
by the use of the growth function $G(z)$, where
$\delta(k,r)=G(r)\delta(k,0)$ (\cite{Pee80} 1980)
\begin{equation}
G(r) \propto {H(r) \over H_0} \int_{z(r)}^\infty dz' (1+z') \left( {H_0
\over H(z')} \right)^3\,.
\end{equation}
Note that in the matter dominated epoch $G \propto a=(1+z)^{-1}$;
it is therefore convenient to define an auxiliary quantity
$F \equiv G/a$.

The cosmological Poisson equation relates the density field to the
fluctuations in the gravitational potential
\begin{equation}
\Phi = {3 \over 2} \Omega_m \left({H_0 \over k}\right)^2
	\left( 1 +3{H_0^2\over k^2}\Omega_K \right)^{-2} F(r)
	\delta(k,0)\,.
\label{eqn:Poisson}
\end{equation}
Likewise, the continuity equation relates the density and velocity
fields via,
\begin{eqnarray}
{\bf v} =  -i \dot G \delta(k,0){ {\bf k} \over k^2 }\,,
\label{eqn:continuity}
\end{eqnarray}
where overdots represent derivatives with respect to radial distance
$\rad$.

For fluctuation spectra and growth rates of interest here,
reionization of the universe
is expected to occur rather late $z_\ri \la 50$ such that the reionized
media is optically thin to Thomson scattering of CMB photons
$\tau \la 1$.
The probability of last
scattering within $d \rad$ of $\rad$ (the visibility function) is
\begin{equation}
g =  \dot \tau e^{-\tau} = X H_0 \tau_H (1+z)^2 e^{-\tau}\,.
\end{equation}
Here 
$\tau(r) = \int_0^{\rad} d\rad \dot\tau$ is the optical depth out to $r$,
$X$ is the ionization fraction,
\begin{equation}
       \tau_H = 0.0691 (1-Y_p)\Omega_b h\,,
\end{equation}
is the optical depth to Thomson
scattering to the Hubble distance today, assuming full
hydrogen ionization with
primordial helium fraction of $Y_p$. 
Note that the ionization
fraction can exceed unity:
$X=(1-3Y_p/4)/(1-Y_p)$  for singly ionized helium,
$X=(1-Y_p/2)/(1-Y_p)$ for fully ionized helium.  We
assume that
\begin{eqnarray}
X(z)& = &1 - {1 \over 2}{\rm erfc}\left(
   {z_{\rm ri}-z \over \sqrt{2} \Delta z} \right)
\end{eqnarray}
such that hydrogen reionizes smoothly but
promptly at $z_{\rm ri}$; some aspects of the
Doppler effect, in particular its power spectrum and to a lesser extent
Doppler-Doppler-OV coupling, are sensitive to the sharpness of this transition.

Although we maintain generality in all derivations, we
illustrate our results with the currently favored $\Lambda$CDM
cosmological model. The parameters for this model
are $\Omega_c=0.30$, $\Omega_b=0.05$, $\Omega_\Lambda=0.65$, $h=0.65$,
$Y_p = 0.24$, $n=1$, and $\delta_H=4.2 \times 10^{-5}$.
This model has mass fluctuations on the $8 h$ Mpc$^{-1}$
scale in accord with the abundance of galaxy clusters
$\sigma_8=0.86$.  A reasonable value here
is important since  the bispectrum is nonlinearly dependent on the
amplitude of the density field. 
We consider reionization redshifts in the range $5 \la z_\ri \la 40$ or
$0.025 \la  \tau \la 0.5$ and assume 
$\Delta z/(1+z)=0.1$.

\subsection{Bispectrum}
\label{sec:bispectrum}

The bispectrum $\bi$
is the spherical harmonic transform of the three-point
correlation function just as the angular power spectrum $C_\ell$
is the transform of the two-point function.
In terms of the multipole moments of the
temperature fluctuation field $T(\hat{\bf n})$,
\begin{equation}
a_{lm} = \int d\bn T(\bn) \Ylmn {}^*(\bn)\,,
\end{equation}
the two point correlation function is given by
\begin{eqnarray}
C(\bn,\bm) &\equiv& \langle T(\bn) T(\bm) \rangle  \nonumber\\
	   &=& \sum_{l_1 m_1 l_2 m_2} \langle \alm{1}^* \alm{2} \rangle
               \Ylmn{}^*(\bn) \Ylmn(\bm)\,.
\label{eqn:twopoint}
\end{eqnarray}
Under the assumption that the temperature field is statistically
isotropic, the correlation is independent of $m$
\begin{eqnarray}
\langle \alm{1}^* \alm{2}\rangle = \deld_{l_1 l_2} \deld_{m_1 m_2}
	C_{l_1}\,,
\end{eqnarray}
and called the angular power spectrum.
Likewise the three point correlation function is given by
\begin{eqnarray}
B(\bn,\bm,\bl) &\equiv& \langle T(\bn)T(\bm)T(\bl) \rangle \\
	       &\equiv&
		\sum 
                \langle \alm{1} \alm{2} \alm{3} \rangle
		\Ylm{1}(\bn) \Ylm{2}(\bm)  \Ylm{3}(\bl)\,,\nonumber
\end{eqnarray}
where the sum is over $(l_1,m_1),(l_2,m_2),(l_3,m_3)$.
Statistical isotropy again allows us
to express the correlation in terms an $m$-independent function,
\begin{eqnarray}
\langle \alm{1} \alm{2} \alm{3} \rangle  = \wjm \bi\,.
\end{eqnarray}
Here the quantity in parentheses is the Wigner-3$j$ symbol
Its orthonormality relation
Eq.~(\ref{eqn:ortho})
implies
\begin{eqnarray}
\bi = \sum_{m_1 m_2 m_3}  \wjm
		\langle \alm{1} \alm{2} \alm{3} \rangle \,.
\label{eqn:bispectrum}
\end{eqnarray}

The angular bispectrum, $\bi$, contains all the information available
in the three-point correlation function.  For example, the skewness,
the collapsed three-point function of \cite{Hinetal95} (1995) and the
equilateral configuration statistic of \cite{Feretal98} (1998) can all
be expressed as linear combinations of the bispectrum
terms (see \cite{Ganetal94} 1994 for explicit expressions).

It is also useful to note its relation to the bispectrum defined on a
small flat section of the sky. In the flat sky approximation, the
spherical polar coordinates $(\theta,\phi)$ are replaced with
radial coordinates on a plane $(r=2\sin\theta/2 \approx \theta,\phi)$.
The Fourier variable conjugate to these coordinates is a 2D vector
${\bf l}$ of length $l$ and azimuthal angle $\phi_l$.  The expansion
coefficients of the Fourier transform of a given ${\bf l}$ is
a weighted sum over $m$ of the spherical harmonic moments of the
same $l$ (\cite{Whietal99} 1999)
\begin{equation}
a({\bf l}) = \sqrt{ 4\pi \over 2l+1} \sum_m i^{-m} a_{l m} e^{im\phi_l}\,,
\end{equation}
so that
\begin{eqnarray}
\left< a^*({\bf l}_1) a({\bf l}_2) \right>
 & = &  {2\pi \over l_1}\deld_{l_1,l_2} C_{l_1} 
		\sum_m e^{im(\phi_{l_1}-\phi_{l_2})}
	\nonumber\\
 &\approx& (2\pi)^2 \deld({\bf l}_1 + {\bf l}_2) C_{l_1}\,.
\end{eqnarray}
Likewise  the 2D bispectrum is defined as
\begin{eqnarray}
\left< a({\bf l}_1) a({\bf l}_2) a({\bf l}_3) \right>
&\equiv& (2\pi)^2 \deld({\bf l}_1+{\bf l}_2 + {\bf l}_3) B({\bf l}_1, {\bf l}_2, {\bf l}_3)
	\\
&\approx&
	{(2\pi)^{3/2} \over (l_1 l_2 l_3)^{1/2}} \bi 
\sum_{m_1,m_2} e^{i m_1 (\phi_{l_1} -\phi_{l_3})}
\nonumber\\
&&\times	
	               e^{i m_2 (\phi_{l_2} -\phi_{l_3})} 
	\wjma{l_1}{l_2}{l_3}{m_1}{m_2}{-m_1-m_2} \,. \nonumber
\end{eqnarray}
The triangle inequality of the Wigner-3$j$ symbol
becomes a triangle equality relating the 2D vectors.  The implication is
that the triplet ($l_1$,$l_2$,$l_3$) can be considered to contribute to
the triangle configuration ${\bf l}_1$,${\bf l}_2$,${\bf l}_3=-{\bf l}_1+
{\bf l}_2$ where the multipole number is taken as the length of the
vector.  

\subsection{Fisher Matrix}
\label{sec:fisher}
To quantify the amount of information contained about a parameter $p_i$
in the bispectrum, we shall define the Fisher matrix ${\bf F}_{ij}$.
The Cramer-Rao inequality (\cite{KenStua69} 1969)
says that the variance of an unbiased
estimator of $p_i$ cannot be less than $({\bf F}^{-1})_{ii}$.
In terms of
the likelihood $L$
of observing bispectrum elements $B_\alpha \equiv \bi$ (arranged
as a data vector) given the true parameters ${\bf p}$ (called
the fiducial model),
the Fisher matrix is defined as
\begin{eqnarray}
{\bf F}_{ij} &\equiv& -\left\langle {\partial^2 \ln  L({\bf B};{\bf p})
	\over
	\partial p_i \partial p_j} \right\rangle\,.
\end{eqnarray}
Under the approximation that the likelihood is Gaussian, this
expression becomes
\begin{eqnarray}
{\bf F}_{ij}     &=& \sum_{\alpha\alpha'}
		{\partial B_\alpha \over \partial p_i}
		({\bf C}^{-1})_{\alpha\alpha'}
		{\partial B_{\alpha'} \over \partial p_j} \,.
\end{eqnarray}
The covariance
matrix between the $\alpha$ and $\alpha'$ bispectrum term is
in general a complicated quantity to calculate (see, \cite{Hea98} 1998).
However there are two simplifying assumptions that make 
this expression tractable.  
Firstly, since the CMB is expected to be nearly Gaussian, the dominant
contribution to the covariance comes from the 6-point function of
the Gaussian field. This can be expressed in terms the
power spectrum of all contributions to the field combined, i.e.
the cosmic signal, detector noise, and residual foregrounds
in the map,
\begin{eqnarray}
C_l^{\rm tot} = C_l + C_l^{\rm noise} + C_l^{\rm foreg}\,.
\label{eqn:cltot}
\end{eqnarray}
Secondly, assuming all-sky coverage\footnote{Even satellite missions will
not have all-sky coverage due to the need to remove galactic contamination.
This will increase the covariance of the estimators but we neglect such
subtleties here.}
the power spectrum covariance
is diagonal in $l,m$ and the covariance then becomes (\cite{Luo94} 1994)
\begin{eqnarray}
{\bf C}_{\alpha\alpha'} &\equiv& \langle {\bi \bip } \rangle \nonumber\\
	&=& C^{\rm tot}_{l_1} C^{\rm tot}_{l_2} C^{\rm tot}_{l_3}
	\Big[
	\deld(123)+\deld(231)+\deld(312) \nonumber\\
        && +\deld(213)+\deld(132)+
		\deld(321) \Big] \,,
\end{eqnarray}
where
\begin{eqnarray}
\deld(abc)=\deld_{l_1 l_a'}
	   \deld_{l_2 l_b'}
	   \deld_{l_3 l_c'} \,.
\end{eqnarray}
Here, we have assumed that $l_1 \ne l_2$, $l_2 \ne l_3$, $l_1 \ne l_3$.
The covariance increases by a factor of 2 for two $l$'s equal
and a factor of 6 for all 3 $l$'s equal.

Under these assumptions, the Fisher matrix reduces to
\begin{eqnarray}
{\bf F}_{ij} &=& \sum_{l_3\ge l_2\ge l_1}
	\sigma_{l_1 l_2 l_3}^{-2}
		{\partial \bi \over \partial p_i}
		{\partial \bi \over \partial p_j} \,,
\end{eqnarray}
where
\begin{eqnarray}
\label{eqn:sigma}
\sigma^2_{l_1 l_2 l_3}
&=&  C_{l_1}^{\rm tot}
	  C_{l_2}^{\rm tot}
	  C_{l_3}^{\rm tot} \\
&& \times \left[ 1 + \deld_{l_1 l_2} + \deld_{l_2 l_3}
+\deld_{l_3 l_1} + 2 \deld_{l_1 l_2} \deld_{l_2 l_3}
\right]\,. \nonumber 
\end{eqnarray}
Note that the covariance between permutations of
($l_1,l_2,l_3$) restricts the sum to ${l_3\ge l_2\ge l_1}$.

Since the signal is expected to be small, in this paper
we will be mainly concerned with the overall
observability of the bispectrum.  Consider the most optimistic
scenario where the form of the bispectrum is considered known
and the only parameter of interest is its amplitude, 
i.e. $\bi = A \bi |_{\rm fid}$ where the true value of $A=1$.
The Fisher matrix tells us that the variance of the measurements
of $A$ is no less than $\sigma^2(A)= ({\bf F}^{-1})_{AA}$ or
\begin{equation}
\chi^2 \equiv {1 \over \sigma^2(A)} =
\sum_{l_3\ge l_2 \ge l_1}
	\frac{\bi^2}{\sigma^2_{l_1 l_2 l_3}}\,,
\label{eqn:chisq}
\end{equation}
which corresponds to the statistic
introduced by \cite{SpeGol99} (1999).  We will often plot the
contribution to $\chi^2$ from a single $l_3$, summed over $l_1$,$l_2$ as
$d\chi^2/dl_3$. 

\subsection{Detector Noise \& Foregrounds}

Detector noise will degrade the sensitivity of a given experiment to the
bispectrum through Eq.~(\ref{eqn:cltot}).
As pointed out by \cite{Kno95} (1995), detector noise in a given
frequency channel $\nu$
can be treated as an additional sky signal with a power spectrum
\begin{equation}
C_l^{\rm noise}(\nu) = w^{-1}(\nu) e^{\theta^2(\nu) \ell(\ell+1)} \,,
\end{equation}
if the experimental beam is Gaussian with width $\theta(\nu)$ in
radians (the full-width-half-maximum is given by $FWHM =
\sqrt{8 \ln 2} \theta(\nu)$).
Here the sensitivity measure $w^{-1}(\nu)$ is defined as the noise
variance per pixel times the pixel area in steradians.
Modern experiments have many
frequency channels with independent pixel noise.  In this
case we inverse variance weight the noise so that
\begin{equation}
{1 \over C_l^{\rm noise}} = \sum_\nu  {1 \over C_l^{\rm noise}(\nu)}
\end{equation}
where the sum is over the different frequency channels.

We will use the MAP and Planck satellite specifications to illustrate
the effects of detector noise (see Table \ref{tab:specs}).
Note that this is somewhat
optimistic since the multifrequency coverage of these experiments
will have to be used to remove foreground contamination.
\cite{Tegetal99} (1999) have shown that given current expectations
for foreground contributions, the increase in $C_l^{\rm tot}$ due
to foreground removal is not expected to exceed $\sim 10\%$.
Of course, the underlying premise that foregrounds add noise as a
Gaussian random field is certainly incorrect at some level.  It may well be that
the bispectrum of the foregrounds is the limiting factor for
observations of the cosmic bispectrum; a  subject which lies beyond
the scope of this paper.  Fig.~\ref{fig:cl} displays the
total power as a sum of the cosmic spectrum and inverse-variance
weighted noise spectrum for the MAP and Planck satellites.  

\begin{table}[tb]\footnotesize
\caption{\label{tab:specs}}
\begin{center}
{\sc CMB Experimental Specifications}
\begin{tabular}{rcccc}
\tableskip\hline\hline\tableskip
Experiment & $\nu$ & FWHM & $10^6 \Delta T/T$ &  \\
\tableskip\hline\tableskip
MAP
& 22 & 56 & 4.1  \\
& 30 & 41 & 5.7  \\
& 40 & 28 & 8.2  \\
& 60 & 21 & 11.0 \\
& 90 & 13 & 18.3 \\
\tableskip\hline\tableskip
Planck
& 30  & 33 & 1.6 \\
& 44  & 23 & 2.4 \\
& 70  & 14 & 3.6 \\
& 100 & 10 & 4.3 \\
& 100 & 10.7 & 1.7 \\
& 143 & 8.0 & 2.0  \\
& 217 & 5.5 & 4.3  \\
& 353 & 5.0 & 14.4 \\
& 545 & 5.0 & 147  \\
& 857 & 5.0 & 6670 \\
\tableskip\hline
\end{tabular}
\end{center}
NOTES.---%
Specifications used for MAP and Planck.
Full width at half maxima (FWHM) of the beams are in arcminutes.
$w^{-1/2} = \Delta T \times {\rm FWHM} \times \pi/10800$.
\end{table}

\subsection{Secondary Anisotropies}
\label{sec:power}

Secondary anisotropies in the CMB are those that are produced well
after recombination at $z \sim 1000$.  They generally fall into two 
broad classes: those arising from gravitational effects and those
arising from Compton scattering. 

\begin{figure}[t]
\centerline{\psfig{file=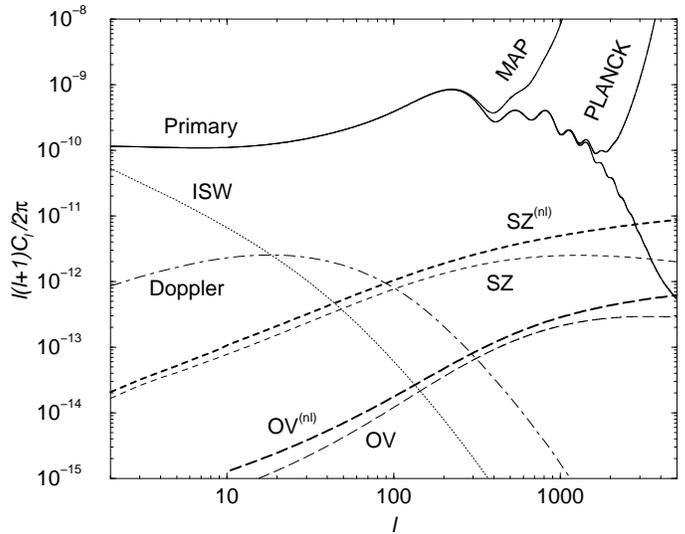,width=3.5in,angle=-90}}
\caption{Power spectrum for the temperature 
anisotropies in the fiducial $\Lambda$CDM model
with $\tau=0.1$ $(\langle z_\ri \rangle = 13\;)$ $\Delta z=0.1(1+z_\ri)$ (see
\S \ref{sec:power} for details).  The curve labeled ``primary'' 
actually includes the small ISW, Doppler, and lensing contributions.  
Note that the predictions for the SZ power spectrum are highly
uncertain and frequency dependent. 
We have also shown the
instrumental noise contribution of MAP and Planck calculated using
parameters in Table~1 
which is important for signal-to-noise calculations.}
\label{fig:cl}
\end{figure}

The gravitational effects
are due to gravitational redshift and lensing.  The 
differential redshift effect from photons climbing in and
out of a time-varying gravitational potential along the line of
sight is called the integrated Sachs-Wolfe 
(ISW; \cite{SacWol67} 1967) effect while 
fluctuations are still linear, and the Rees-Sciama 
(\cite{ReeSci68} 1968) effect for the second-order and nonlinear contributions. 
The ISW effect is important for low matter density universes
$\Omega_m < 1$, where the gravitational potentials decay at low redshift,
and contributes anisotropies on and above the scale of the horizon at the
time of decay (see Fig.~\ref{fig:cl} 
calculated from Eq.~[\ref{eqn:clsecondary}]).
The Rees-Sciama effect is small in adiabatic 
CDM models (\cite{Sel96} 1996) and we will not consider it further here.

Gravitational lensing of the photons by the intervening large-scale structure 
both redistributes power in multipole space and enhances it due power 
in the density perturbations.  The most effective structures
for lensing lie half way between the surface of recombination and
the observer in comoving angular diameter distance.  In the
fiducial $\Lambda$CDM cosmology, this is at $z \sim 3.3$, 
but the growth of structure skews this to
somewhat lower redshifts.  In general, the efficiency of lensing is 
described by a broad bell shaped function between the source and the
observer.  The curve labeled ``primary'' in Fig.~\ref{fig:cl} includes
the gravitational lensing effect on the power spectrum 
as calculated by the CMBFAST code
(\cite{SelZal96} 1996 and \cite{ZalSel97} 1997); lensing is responsible
for the power-law tail of anisotropies for $l>4000$ here.  

Rescattering of the photons during the reionized epoch can both generate
and erase anisotropies.   The primary anisotropies are reduced as $\exp(-\tau)$
by scattering on small scales $l \ga 50$.  Density fluctuations can lead
to optical depth variations and hence a patchy erasure of fluctuations.
For the optically thin conditions considered here, this is a negligible effect.

The bulk flow of the electrons that scatter the CMB photons leads to  
a Doppler effect.  Its effect on the power spectrum  peaks around the 
horizon at the scattering event projected on the sky 
today (see Fig.~\ref{fig:cl}).  
The contributions are strongly dependent on the optical
depth in the reionized epoch; we take $\tau=0.1$ as our fiducial model
but explore optical depths up to $\tau=0.5$.
On scales smaller than the horizon at scattering, the contributions are
mainly canceled as photons scatter against the crests and troughs of
the perturbation.  As a result, the Doppler effect is moderately sensitive
to how rapidly the universe reionizes since contributions from a sharp
surface of reionization do not cancel.   Following current thinking
that the universe reionized promptly (see \cite{HaiKno99} 1999 for a review), 
we take $\Delta z = 0.1(1+z_\ri)$ in Fig.~\ref{fig:cl}.

The Sunyaev-Zel'dovich (SZ; \cite{SunZel80} 1980) effect arises 
from the  inverse-Compton scattering of CMB photons by hot electrons 
along the line of sight. This effect has now been directly imaged 
towards massive galaxy clusters (e.g., \cite{Car96} 1996), 
where temperature of the scattering medium  can reach as high as 
10 keV producing temperature changes in the CMB of order 1 mK. 
Here, we are interested in the SZ effect produced by large-scale 
structure in the general intergalactic medium (IGM), where the exact
calculation on its significance as a secondary anisotropy is
not directly possible due to the unknown distribution and 
clustering properties of baryonic gas and its temperature structure.   
Under certain simplifying, albeit largely untested assumptions 
detailed in \S~\ref{sec:SZlensing},  its power spectrum is given
in Fig.~\ref{fig:cl} for both linear and nonlinear contributions.
Given the untested assumptions, these spectra should be taken as provisional, 
even in their order of magnitude.  Note that the SZ effect also bears 
a spectral signature that differs from the other effects.  
We have assumed observations in the Rayleigh-Jeans
regime of the spectrum; an experiment such as Planck with sensitivity
beyond the peak of the spectrum can separate out these contributions
based on the spectral signature.   

The Ostriker-Vishniac effect arises from the second-order
modulation of the Doppler effect by density fluctuations
(\cite{OstVis86a} 1986a; \cite{Vis87} 1987).  Its nonlinear
analogue is the kinetic SZ effect from large-scale structure 
(\cite{Hu99} 1999); we denote it OV$^{(\rm nl)}$ here to avoid 
confusion with the thermal SZ effect.  Due to its density weighting,
the OV effect peaks at small scales: arcminutes for $\Lambda$CDM.  
For a fully ionized universe,
 contributions are broadly distributed in redshift so that the 
power spectra are moderately dependent on the optical depth $\tau$, 
which we generally take to be $0.1$.  

Variations in the ionization fraction of the gas around the epoch of
reionization can also modulate the Doppler effect (\cite{Aghetal96} 1996;
\cite{GruHu98} 1998; \cite{KnoScoDod98} 1998).  This effect depends on
the detailed physical processes occurring at reionization though it probably
contributes substantially to the sub-arcminute scale anisotropy.  
Because of the small scale of the effect and its unknown relationship
to the density field and hence the other secondary effects, we do not
consider its contribution to the power spectrum or bispectrum here.

In summary, for $\Lambda$CDM cosmologies with optical depths $\tau \sim 0.1$,
the ISW contributions dominate at low $l$-values, 
the Doppler contributions at intermediate $l$-values, 
and the SZ contributions at small
scales.  Given the crudeness of the approximation for the SZ effect, 
one cannot rule out the possibility that OV contributions dominate at 
high-$l$.  
The power spectra of these effects imply that the intrinsic contributions
of the ISW, Doppler and SZ effects are comparable but appear at different
angular scales and arise from different redshifts.  It is for this reason
that we shall consider the bispectrum contributions from these effects
in the following sections.

\section{Lensing Effects: Derivation}
\label{sec:lensing}

The general derivation of CMB bispectrum due to the coupling of
lensing with secondary
anisotropies, originally considered by \cite{GolSpe99} (1999), is reviewed
in \S \ref{sec:generallensing}.  It is applied to
Doppler secondary anisotropies in \S \ref{sec:dopplerlensing}.
In \S~3.3 and 3.4 for comparison, we briefly revisit 
the coupling between gravitational lensing and ISW and SZ effects
respectively studied by \cite{GolSpe99} (1999).

\subsection{General Considerations}
\label{sec:generallensing}

Large-scale structure deflects CMB photons in transit from the
last scattering surface.  These structures also give rise
to secondary anisotropies.   The result is a correlation between
the temperature fluctuations and deflection angles.
This effect cannot be seen in the two point function since
gravitational lensing preserves surface brightness: deflections
only alter the temperature field on the sky in the presence of
intrinsic, primary, anisotropies in the unlensed distribution.
The lowest order contribution thus comes from the three-point
function or bispectrum.

In weak gravitational lensing, the deflection angle on the sky is
given by the angular gradient of the lensing
potential which is itself a projection of the gravitational
potential (see e.g. \cite{Kai92} 1992),
\begin{eqnarray}
\Theta(\bm)
&=&
- 2 \int_0^{\rad_0} d\rad \frac{\da(\rad_0-\rad)}{\da(\rad)\da(\rad_0)}
		\Phi (\rad,\hat{{\bf m}}\rad ) \,.
\end{eqnarray}
This quantity is simply related to the more familiar
convergence
\begin{eqnarray}
\kappa(\bm) & = &{1 \over 2} \nabla^2 \Theta(\bm) \\
            & = &-\int_0^{\rad_0} d\rad \frac{\da(\rad)\da(\rad_0-\rad)}{\da(\rad_0)}
\nabla_{\perp}^2 \Phi (\rad ,\hat{{\bf m}}\rad) \, , \nonumber\\
\nonumber
\end{eqnarray}
where note that the 2D Laplacian operating on $\Phi$ is
a spatial and not an angular Laplacian.
The two terms $\kappa$ and $\Theta$ contain superficial differences
in their radial and wavenumber weights which we shall see cancel
in the appropriate Limber approximation.  In particular,
their spherical harmonic moments are simply proportional
\begin{eqnarray}
\Theta_{l m} &=&
             -{2 \over l(l+1)} \kappa_{l m} =
		 \int d {\bn} \Ylmn{}^*(\bn) \Theta(\bn) \nonumber\\
             &=& i^l \int {d^3 {\bf k}\over 2\pi^2} \delta({\bf k})
		\Ylmn{}^* (\bk) I_\ell^{\rm len}(k)
\label{eqn:GSSZequiv}
\end{eqnarray}
with
\begin{eqnarray}
I_\ell^{\rm len}(k)& =&
		\int_0^{\rad_0} d\rad W^{\rm len}(k,r)
		 j_l(k\rad)  \,,\nonumber\\
W^{\rm len}(k,r)& =&
		-3 \Omega_m \left({H_0 \over k}\right)^2
		F(r) {\da(\rad_0 - \rad) \over
		\da(\rad)\da(\rad_0)}\,.
\label{eqn:lensint}
\end{eqnarray}
Here, we have used the Rayleigh expansion of a plane wave
\begin{equation}
e^{i{\bf k}\cdot \hat{\bf n}\rad}=
4\pi\sum_{lm}i^lj_l(k\rad)Y_l^{m \ast}(\bk)
\Ylmn(\bn)\,,
\label{eqn:Rayleigh}
\end{equation}
and the fact that $\nabla^2 \Ylmn = -l(l+1) \Ylmn$.  In an open
universe, one simply replaces the spherical Bessel functions with
ultraspherical Bessel functions in expressions such as
Eq. (\ref{eqn:lensint}).  Since this
is the case, we present the derivation in flat space.

Since expressions of the type in Eq.~(\ref{eqn:lensint}) frequently
occur in the following calculations, it is useful to note that it may
be evaluated efficiently with a temporal 
version of the Limber approximation (\cite{Lim54} 1954)
called the weak coupling approximation (\cite{HuWhi96} 1996),
\begin{eqnarray}
I_l^{\rm X}(k) &\equiv &
		\int_0^{\rad_0}
		 d\rad W^{\rm X}(k,r) j_l(k\rad)  \nonumber\\
	  &\approx&
		W^{\rm X}(k,l/k) \int_0^\infty d\rad j_l(k\rad)  \qquad
		(k {W \over \dot W} \gg 1) 	\nonumber\\
	  &= &
		W^{\rm X}(k,l/k) {\sqrt{\pi} \over 2 k}
		     {\Gamma[(l+1)/2]\over \Gamma[(l+2)/2]}\,,
\label{weakcoupling}
\end{eqnarray}
and note that the ratio of gamma functions goes to $\sqrt{2/l}$ for
$l\gg 1$.  For the open universe generalization of this result, see
\cite{Hu99} (1999).   We employ this approximation for sufficiently
high $l$ values (usually $l > 200$), where the difference between full integral and
the approximation is sufficiently low $(<1\%)$.

Assuming a relation between the multipole moments and $I^{\rm X}_l$
as in Eq.~(\ref{eqn:GSSZequiv}), we find that the cross-correlation
power between X and Y then becomes
\begin{eqnarray}
\left< a_{lm}^{\rm X*} a_{lm}^{\rm Y} \right>
&=& 4\pi \int {d k \over k} \Delta^2(k) I_l^{\rm X}(k) I_l^{\rm Y}(k) \\
&\approx& { 2 \pi^2 \over l^3 } \int_0^{\rad_0} d\rad\, \da
	W^{\rm X}({l \over \da},r) W^{\rm Y}({l \over\da},r)
	 \Delta^2({l \over\da}) \nonumber\,,
\label{eqn:powerform}
\end{eqnarray}
where the  approximation involves a change in variables $\da=l/k$ and
we have restored generality for open geometries where $r \ne \da$
(see \cite{Hu99} 1999).
Note that in the Limber approximation where $\da$ is interchangeable
with $l/k$, the superficial difference in the weights for $\kappa$
and $\Theta$ disappears.

The quantity of interest is the correlation between the deflection
potential and secondary anisotropies
\begin{equation}
T^{\rm S}(\bn) = \sum \almn^{\rm S} \Ylmn(\bn),
\end{equation}
which becomes
\begin{eqnarray}
\langle \Theta(\hat{\bf n})T^{\rm S}(\hat{\bf m})\rangle&=&
\sum_{l m}
	\left< \Theta_{l m}^* \almn^{\rm S} \right>
           \Ylmn{}^*(\bn) \Ylmn (\bm)\,.
\end{eqnarray}
Again statistical isotropy guarantees that we may write the correlation
as
\begin{eqnarray}
	\left< \Theta_{l m}^* \almn^{\rm S} \right> & \equiv & b_l^{\rm S}
	 \equiv  {-2\over l(l+1)} C_l^{T \kappa} \,, \nonumber\\
\label{eqn:bl}
	 &=& 4\pi \int {dk \over k} \Delta^2(k) I_l^{\rm S}(k) 
		I_l^{\rm len}(k) \,, \\ 
	 &\approx&
        { 2 \pi^2 \over l^3 } \int_0^{\rad_0} d\rad\, \da
	W^{\rm S}({l \over \da},r) W^{\rm len}({l \over\da},r) 
	\Delta^2({ l \over \da}) \,, \nonumber
\end{eqnarray}
where we have used equation~(\ref{eqn:GSSZequiv}) to
relate the power spectrum
$b_l^{\rm S}$ defined by \cite{GolSpe99} (1999) and the $\kappa$-secondary
cross power spectrum defined by \cite{SelZal99} (1999).  The last line
represents the Limber approximation and we have assumed that 
the secondary anisotropies are linearly related to the density field
projected along the line of sight,
\begin{eqnarray}
a^{\rm S}_{lm} &=& i^l \int \frac{d^3\veck}{2 \pi^2}
\delta(\veck)  I_l^{\rm S}(k) \Ylmn(\hat{\veck}) \, , \nonumber\\
I_l^{\rm S}(k) &=& \int d\rad  W^{\rm S}(k,\rad)j_{l}(k\rad) \, .
\label{eqn:secondaryform}
\end{eqnarray}
Note that the power spectrum of the secondary effect is then 
given by
\begin{equation}
C_l^\se = 4\pi \int {dk \over k} \Delta^2(k)
		I_l^\se(k) I_l^\se(k) \,.
\label{eqn:clsecondary}
\end{equation}
We will discuss explicit forms for $W^{\rm S}$ for specific 
secondary effects in the following sections.

Unfortunately, the secondary-lensing correlation is not directly observable.
As pointed out by \cite{GolSpe99} (1999), it does however have
an effect on the bispectrum which is in principle observable.
The lensed temperature fluctuation in a given direction is the sum of
the primary fluctuation in a different direction plus the secondary
anisotropy
\begin{eqnarray}
T(\bn) &=& T^{\rm P}(\bn + \nabla \Theta) + T^{\rm S}(\bn)  \\
       &\approx&
        \sum_{lm} \Big[ (\almn^{\rm P}+\almn^{\rm S}
		  )\Ylmn(\hat{\bf n})
                  +\almn^{\rm P}  \nonumber\\
       &&\times
	\nabla\Theta(\hat{\bf n})\cdot\nabla \Ylmn(\hat{\bf n}) \Big]
\, ,
\nonumber
\end{eqnarray}
or
\begin{eqnarray}
\almn &=& \almn^{\rm P} + \almn^{\rm S}
	+ \sum_{l'm'}
	a_{l' m'}^{\rm P}\nonumber\\
&&\times
	\int d \bn \Ylmn{}^* (\bn)
	\nabla\Theta(\hat{\bf n})\cdot
	\nabla Y_{l'}^{m'}(\hat{\bf n})
	\,.
\end{eqnarray}
Utilizing the definition of the bispectrum 
in Eq.~(\ref{eqn:bispectrum}), we obtain
\begin{eqnarray}
\bi &=& \sum_{m_1 m_2 m_3} \wjm
\nonumber \\
&&\times \int d\hat{\bf m} d\hat{\bf n}
\Ylm{2}{}^*(\bm)
\Ylm{3}{}^*(\bn) C_{l_1}
\nonumber \\
&&\times
\nabla \Ylm{1}{}^*(\bm)
\cdot
\langle \nabla\Theta(\bm)
T^{\rm S}(\hat{\bf n}) \rangle  + {\rm Perm.}
\end{eqnarray}
where the five permutations are with respect to the ordering of
$(l_1,l_2,l_3)$.

Integrating by parts and simplifying further following
\cite{GolSpe99} (1999) leads to a
bispectrum of the form:
\begin{eqnarray}
&& \bi = -\wj
\sqrt{ \frac{(2l_1 +1)(2 l_2+1)(2 l_3+1)}{4 \pi}}
\nonumber \\
&\times&
\left[\frac{l_2(l_2+1)-l_1(l_1+1)-l_3(l_3+1)}{2} C_{l_1}b^{\rm S}_{l_3}+ {\rm Perm.}\right]\,, \nonumber \\
\end{eqnarray}
where we have employed Eq.~(\ref{eqn:harmonicsproduct}) to perform the angular
integration.

\subsection{Doppler-Lensing Effect}
\label{sec:dopplerlensing}

The Doppler effect generates temperature fluctuations as
\begin{equation}
T^\dop(\bn) = \int_0^{\rad_0} d\rad g(r) \bn \cdot {\bf v}(\rad,\bn \rad)\,.
\end{equation}
With the help of Eq.~(\ref{eqn:continuity}), 
\begin{equation}
\hat{\bf n} \cdot \veck = \sum_{m} \frac{4\pi}{3} k
Y_1^{m}(\hat{\bf n}) Y_1^{m\ast}(\hat{\veck}) \,,
\end{equation}
and the Clebsh-Gordan coefficients for the addition of
angular momenta with $l=1$,
we can write the multipole moments as
\begin{eqnarray}
\almn^\dop &=&
	-4\pi i^l
	\int \frac{d^3\veck}{(2\pi)^3} {\delta({\bf k}) \over k}
	\Ylmn{}^*(\bk) \nonumber \\
&\times&	\int_0^{\rad_0}  d\rad
	g \dot G j_{l}'(k\rad) \,.
\label{eqn:dopplersource}
\end{eqnarray}
Through integration by parts, this expression can be brought
into the standard form of Eq.~(\ref{eqn:secondaryform}) with
\begin{eqnarray}
W^{\rm dop}(k,r)&=& {1 \over k^2} (\dot g \dot G + g \ddot G)\,,
\end{eqnarray}
through integration by parts.

Employing Eq. (\ref{eqn:bl}) for the lensing correlation,
we obtain
\begin{eqnarray}
b_l^{\rm dop}
&\approx&
	-{6 \pi^2 \over l^7} \Omega_m H_0^2 \int_0^{\rad_0}d\rad \da^4
      (\dot g \dot G + g \ddot G) \nonumber\\
&&\times	F(r)
	\frac{\da(\rad_0-\rad)}{\da(\rad_0)}
	\Delta^2(l/\da),
\label{eqn:dopplerlensing}
\end{eqnarray}
in the Limber approximation for $l \ga 200$.

\begin{figure}[t]
\centerline{\psfig{file=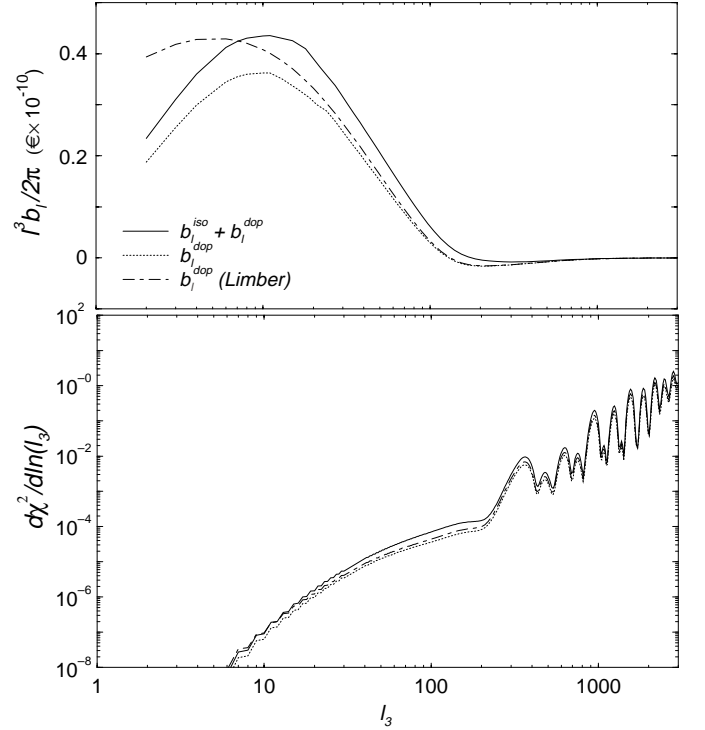,width=3.5in,angle=0}}
\caption{The Doppler-lensing effect.  
Shown are the combined Doppler
and double scattering effects,
({\it solid} line), the Doppler effect ({\it dotted line}),
and the Limber approximation to the Doppler effect 
({\it dot-dashed} line). 
At sufficiently high $l$, the difference between these three
treatments can be ignored for most practical purposes.
{\it Top panel---} The correlation power spectrum.
{\it Bottom panel---} Contribution to $\chi^2$ per log interval in $l_3$.}
\label{fig:bldoppler}
\end{figure}

\subsection{Double-Scattering-Lensing Effect}
\label{sec:doublelensing}

As pointed out by \cite{Kai84} (1984), double scattering effects must
also be considered since the single scattering contributions are canceled
along the line of sight at short wavelengths.
Equation~(\ref{eqn:dopplersource}) implies
the Doppler effect produces an isotropic temperature fluctuation in
the photons which subsequently last scatter into the line of sight,
\begin{eqnarray}
a_{00}^{\rm dop}(k) Y_0^0
& \equiv &
k^{-2} S^{\rm iso}(k,\rad)\delta({\bf k})
\nonumber\\
		       & = & {\delta({\bf k}) \over k}
				\int_{\rad}^{\rad_0} d\rad'
				g \dot G j_1[k(\rad -\rad')] \,,
\end{eqnarray}
where $g$ is to be interpreted as the visibility function 
for an observer at $\rad$,
i.e. $g = \dot\tau \exp[-\tau(\rad')+\tau(\rad)]$.
For scales that are much smaller than the width of the visibility
function, i.e. approximately $k(\rad_0-\rad)\gg 1$,
one can take $g\dot G$ out of the
integral leaving
\begin{equation}
S^{\rm iso}(k,\rad) \approx \dot\tau \dot G \,.
\end{equation}
The contribution to the anisotropy today takes the
standard form of Eq.~(\ref{eqn:secondaryform}) 
\begin{eqnarray}
W^{\rm iso}(k,r) &=& k^{-2} g(r) S^{\rm iso}(k,\rad)\,.
\end{eqnarray}
In the Limber limit, Eq.~(\ref{eqn:bl})
can be simplified as
\begin{eqnarray}
b_l^{\rm iso} 
&\approx&
	-{6 \pi^2 \over l^7} \Omega_m H_0^2 \int_0^{\rad_0}d\rad \da^4
      (g \dot \tau \dot G) \nonumber\\
&&\times	F(r)
	\frac{\da(\rad_0-\rad)}{\da(\rad_0)}
	\Delta^2(l/\da),
\end{eqnarray}
again valid for $\ell \ga 200$.

\subsection{ISW-Lensing Effect}
\label{sec:ISWlensing}

The integrated Sachs-Wolfe effect (\cite{SacWol67} 1967) results from the
late time decay of gravitational potential fluctuations. The resulting
temperature fluctuations in the CMB can be written as
\begin{equation}
T^\isw(\bn) = -2 \int_0^{\rad_0} d\rad \dot{\Phi}(\rad,\bn \rad) \, .
\end{equation}
Using the Poisson equation (Eq.~[\ref{eqn:Poisson}]), we can then 
bring the contributions into the standard form of Eq.~(\ref{eqn:secondaryform})
\begin{eqnarray}
W^\isw(k) = -3\Omega_m \left( {H_0 \over k} \right)^2
	 \dot F(r) \, .
\label{eqn:iswsource}
\end{eqnarray}

The two-point function produced between gravitational lensing angular
deflections and ISW effect can now be written as in the 
Limber limit 
as (\cite{GolSpe99} 1999)\footnote{Note that our Limber 
approximated correlation between lensing and ISW
effect is factor 2 lower than Eq.~(14) of \cite{GolSpe99} (1999); the
same equation also contains an additional misprint
with respect to weighing by their $\tau$ factors.}
\begin{eqnarray}
b_l^\isw 
& \approx & \frac{18 \pi^2}{l^7} {\Omega_m^2 H_0^4} \int_0^{\rad_0} d\rad
      \dot{F} F \da^4 \frac{\da(\rad_0-\rad)}{\da(\rad_0)}
      \Delta^2(l/\da) \, . \nonumber \\
\label{eqn:iswlimber}
\end{eqnarray}
For calculational purposes, we employ numerical integration to 
$l$ of 200 and use
Limber approximated formulation thereafter.

\begin{figure}[t]
\centerline{\psfig{file=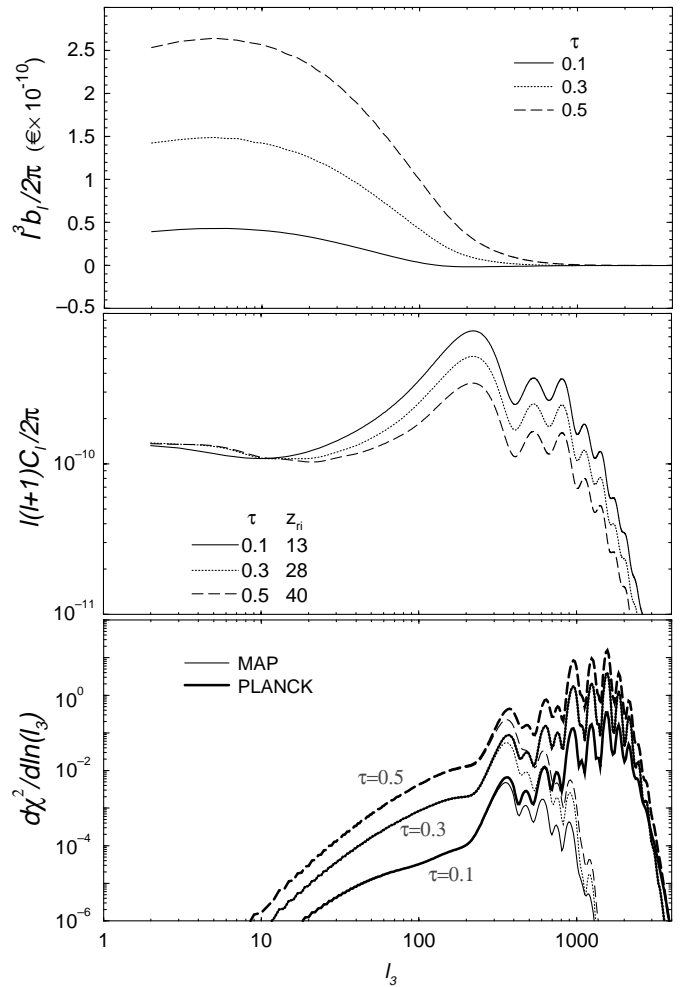,width=3.5in,angle=0}}
\caption{Dependence of the Doppler-lensing effect 
on the ionization optical depth
for $\tau$ of 0.1, 0.3 and 0.5 with $\Delta z = 0.1(1+z_\ri)$.
{\it Top panel---} The $b_l$ term for the Doppler-lensing
effect.
{\it Middle panel---} CMB power spectrum used in the
noise calculation.
{\it Lower panel---}
The contribution to $\chi^2$ per log interval in $l_3$ with MAP and Planck
detector noise included.
}
\label{fig:dopplermp}
\end{figure}

\subsection{SZ-Lensing Effect}
\label{sec:SZlensing}

The SZ effect leads to an effective temperature fluctuation in the
Rayleigh-Jeans part of the CMB given by the integrated pressure fluctuation along 
the line of sight:
\begin{equation}
T^\sz(\bn) = -2 \int_0^{\rad_0} d\rad\, g \left( { k_B T_e \over  m_e c^2} \right)
	\left[\delta_{\rm gas} + {\delta T_e \over T_e}\right](\rad,\bn \rad) \, ,
\end{equation}
where $k_B$ is the Boltzmann constant.
Unfortunately, the clustering properties of the gas and its temperature
structure is uncertain even with state of the art hydrodynamic simulations.

To obtain an order of magnitude estimate of the effect, we follow
\cite{GolSpe99} (1999) in making several simplifying but largely untested
assumptions.
We assume that the gas is a
biased tracer of the dark matter density $\delta_{\rm gas} = b_{\rm gas} \delta$
and that the IGM temperature
distribution varies as
$T_e(r) =  a(r) T_{\rm e0}$.  Finally we ignore temperature fluctuations
in the gas.   

The effect can then be expressed 
as a weighted projection of the
density field of the form in
Eq.~(\ref{eqn:secondaryform})
\begin{eqnarray}
W^\sz(k) & = & -2 A^\sz
	g a G  \, ,
\end{eqnarray}
with a normalization given by
\begin{eqnarray}
A^\sz    & = & b_{\rm gas} \frac{k_B T_{\rm e0}}{m_e c^2} \nonumber\\
	 & = & 0.00783 \left( {b_{\rm gas} \over 4} \right) 
		       \left( {T_{\rm e0} \over 1{\rm keV}} \right)\,.
\end{eqnarray}
For illustration purposes, we adopt $b_{\rm gas}=4$ and $T_{\rm e0}=1$keV
throughout.

The cross power between lensing and SZ can be written
using Eq.~(\ref{eqn:bl}) 
\begin{eqnarray}
b_l^\sz 
        &\approx&
\frac{12 \pi^2}{l^5} \Omega_m H_0^2 A^\sz 
\int d\rad G^2\da^2 g 
      \frac{\da(\rad_0-\rad)}{\da({\rad_0})} \Delta^2\left({l
\over\da}\right) \,. \nonumber \\ 
\end{eqnarray}
Since the SZ-lensing results in a correlation that peaks at $l \sim 100$
(see, \cite{GolSpe99} 1999), with no significant contribution at
low $l$s, the Limber approximation may be used to calculate the whole 
effect. 

\begin{figure}[t]
\centerline{\psfig{file=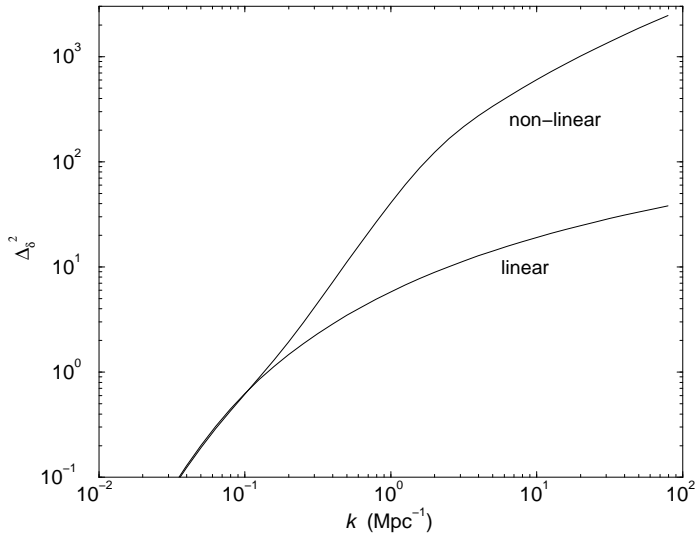,width=3.6in,angle=-90}}
\caption{Linear and nonlinear density power spectra for the dark matter
under the Peacock \& Dodds (1996) scaling approximation for our 
fiducial $\Lambda$CDM cosmological model evaluated at the present.}
\label{fig:deltanl}
\end{figure}

Both SZ effect and gravitational lensing angular deflection potentials 
will be enhanced by nonlinear fluctuations
in the density field.  To get a qualitative understanding of the effects, assume that the gas density 
and potential fluctuations continue to track the dark matter in the nonlinear regime.  
The nonlinear evolution of the dark matter density distribution
has been  well-studied with N-body simulations for
adiabatic CDM cosmological models of interest here. 
We employ the scaling relation of  \cite{PeaDod96} (1996) 
to obtain the nonlinear power spectrum today (see Fig.~\ref{fig:deltanl})
as well as its evolution in time.
In Fig.~\ref{fig:cl}, the curve labeled ``SZ$^{\rm (nl)}$'' is the 
power spectrum formed by the SZ effect using  
the non-linear matter density power spectrum. 
The nonlinear effects generally increases the power due to SZ
temperature fluctuations by a factor of 3 to 4 when $l$ is in the interested
range of 1000 to 5000. 

With the introduction of the non-linear power spectrum, we can
write the cross-correlation between lensing and SZ as
\begin{eqnarray}
b_l^{\rm SZ(nl)} 
        &\approx&
\frac{12 \pi^2}{l^5} \Omega_m H_0^2 A^\sz \nonumber \\
&\times& \int d\rad \da^2 g 
      \frac{\da(\rad_0-\rad)}{\da({\rad_0})} \Delta^{2{\rm (nl)}}\left({l \over\da},\rad\right) \,. 
\end{eqnarray}
Since the nonlinear power spectrum no longer grows as $G^2$, it must
be evaluated along the line of sight at the corresponding lookback time.

\section{Lensing Effects: Results}
\label{sec:lensingresults}

\begin{figure}[t]
\centerline{\psfig{file=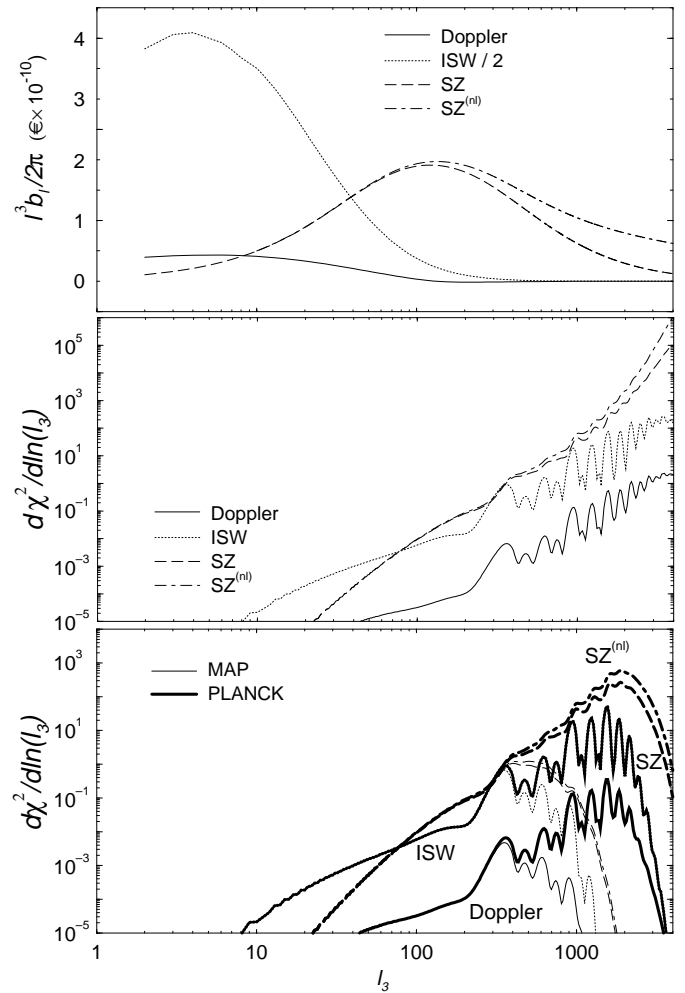,width=3.5in,angle=0}}
\caption{Comparison of various lensing effects
for our fiducial $\Lambda$CDM model
and with $\tau=0.1$ and $\Delta z=0.1 (1+z_\ri)$. 
{\it Top panel---} The power spectrum of the correlation. 
{\it Middle panel---} Contributions to $\chi^2$ per log interval in
$l_3$, assuming cosmic variance only. {\it Lower panel---} The same adding in detector
noise for MAP and Planck.}
\label{fig:dopplerszisw}
\end{figure}

\subsection{Doppler-Lensing Bispectrum}

The contribution to the bispectrum from the coupling of lensing
to the Doppler effect 
is encapsulated in the power spectrum of
the correlation $b_l^\dop$ (see Eq.~[\ref{eqn:dopplerlensing}]).  
Figure~\ref{fig:bldoppler} (top)
shows its value for the Doppler effect in 
our fiducial $\Lambda$CDM $\tau=0.1$ cosmology 
assuming cosmic variance only $C_l^{\rm tot}=C_l$
(see Eq.~[\ref{eqn:cltot}]).
The zero crossing at $l\sim 100$ is 
due to the competition between contributions
on the reionization surface and those from the intermediate
redshifts. 
Likewise the
Limber approximation to the integral also becomes excellent 
beyond $l = 200$.
The inclusion of double scattering contributions to the Doppler effect
only makes a minor difference for $\tau \sim 0.1$.

We also show the contribution to the overall signal to noise squared
($\chi^2$) per logarithmic interval in $l_3$ in Fig.~\ref{fig:bldoppler} (bottom).
The minor differences in the $\chi^2$ contributions at high $l_3$ 
between the Limber-approximated  and Doppler-only spectra from the full
calculation are primarily due to the differences of the $b_l$ term at low
$l$ values.  Low-$l$ values in $b_l$ contribute to high $l_3$ values
in $\chi^2$ since the latter is a sum over $l_1 \le l_2 \le l_3$. 
The structure in the $\chi^2$ plot and the rapid increase of its 
values around $l \sim 1000$ arises in large part from the structure
of the primary $C_l$ itself which determines the Gaussian noise per mode.

Since the signal is dominated by the smallest angular scales available
in the measurements, it is important to include the experimental beam
and instrumental noise contributions.
In Fig.~\ref{fig:dopplermp},  we show the signal-to-noise ratio
per mode of $l_3$ assuming MAP and Planck noise (see Table 1).
Here, we also show the affect of varying the optical depths from 0.1 to 0.5.
Note that because
reionization decreases the power spectrum of the anisotropies
as $e^{-2\tau}$, the Gaussian noise in the bispectrum is also
decreased with increasing optical depth.
We have assumed a reionization width of $\Delta z= 0.1(1+z_\ri)$ 
here, but the lensing
correlation is not sensitive to this parameter.  
The lensing efficiency is a broad bell-shaped function 
in angular diameter distance
and does not correlate well with contributions localized near the
distant surface of reionization.

As shown in Fig.~\ref{fig:dopplermp},  
the bispectrum mode contribution to $\chi^2$ for
any $l_3$  due to coupling between
reionized Doppler effect and gravitational lensing angular deflections
is generally less than $10^{-2}$, unless the optical depth to
reionization is greater than what is currently inferred from
the observed power spectrum of anisotropies ($\tau < 0.5$; 
\cite{Grietal99} 1999; see, \cite{HaiKno99} 1999 for a recent review).  
Adding all the $l_3$ mode contributions, the total $\chi^2$ value is
generally at a level below where one can expect its detection in 
the upcoming satellite data for reasonable optical depths ($\tau \la 0.3$).

\begin{figure}
\centerline{\psfig{file=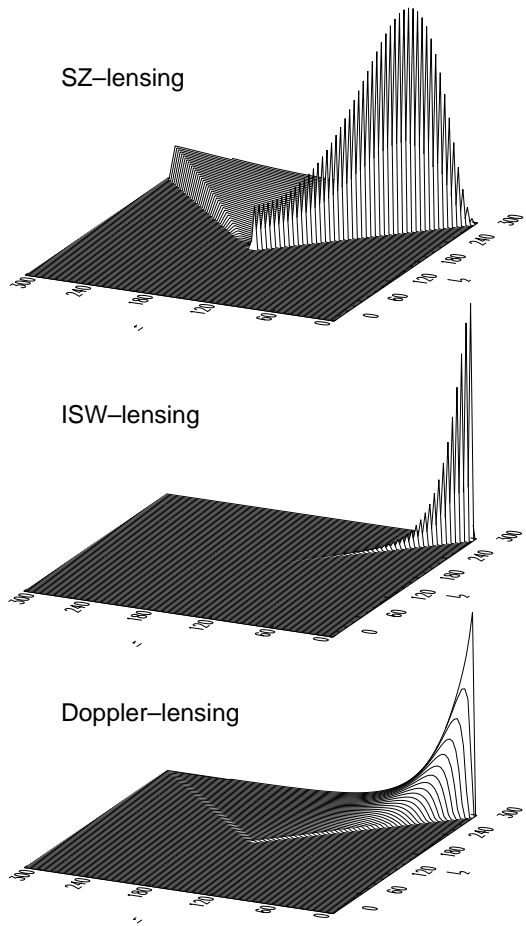,width=2.7in,angle=0}}
\caption{Configuration dependence of the secondary-lensing signal.
Plotted is the contributions to $\chi^2$ of from single mode $l_3=300$
as a function of $l_2$ and $l_1$.  
The configuration shapes are determined by 
both the power spectra of the cross-correlation ($b_l$ terms; see
Fig.~\ref{fig:dopplerszisw}) and the CMB power spectrum.
The spiky behavior in these plots is an artifact of the plotting near
the sharp crest introduced by the Wigner-$3j$ symbol (see 
Fig.~\ref{fig:wig3j}).}

\label{fig:lensconfiguration}
\end{figure}

\subsection{Comparison of Lensing Effects}

In Fig.~\ref{fig:dopplerszisw}, we compare the relative 
contribution to the CMB bispectrum from
all the effects calculated. 
Both ISW and Doppler effects contribute at low $l$ values to the
correlation $l^3 b_l/2\pi$, while SZ effect to $l \sim 100$. 
The Doppler-lensing coupling is
suppressed by a factor of $\sim$ 8 when compared to ISW-lensing
coupling,  which is in large part due to the inefficiency with which
high redshift structures lens the CMB. 
As shown, the SZ-lensing coupling and ISW-lensing coupling dominate
the Doppler-lensing effect for  optical depths $\tau \la 0.5$. 
The nonlinear growth of fluctuations increases the SZ-lensing
cross-correlation when $l > 100$ when compared to pure linear
effects with a modest enhancement factor of $\sim$ 5 at $l \sim 5000$. Such
an enhancement in the SZ-lensing cross-correlation power spectrum is consistent
with increases in both SZ and lensing individual power spectra due to
nonlinear effects.

If untested assumptions on the SZ thermal effect were found to be false and the
SZ contribution to be lower than currently presumed, than most of the
contribution, at least in a low $\Omega_m$ flat cosmological model, would
come from coupling between gravitational lensing and the ISW
effect. 
However, even if this turns out to be the case,
one may be able to use the spectral signature of the SZ effect
to separate its contribution.
If SZ-lensing coupling is clearly detected, it would
also support the current notion that most of baryons are present in
the IGM as form of a gas at temperature of $\sim$ 1 keV. As suggested
by \cite{GolSpe99} (1999), this would certainly solve the
``missing baryons'' problem (\cite{CenOst99} 1999). 
In the meantime, improvements of our knowledge of the physical nature of 
the baryonic gas is clearly needed and can 
come from both observations as well as numerical
simulations (e.g., \cite{daS99} 1999).

Another obstacle for the unambiguous detection of the SZ, ISW and
Doppler lensing effects is the ability to distinguish their
bispectrum contribution from those of other physical mechanisms.  
The $\chi^2$ statistic of \cite{GolSpe99} (1999) lumps all the information
on each effect
into one number and makes such an assessment difficult.  
Once all the effects are known the question of separability can be
addressed with the Fisher matrix techniques of \S \ref{sec:fisher}.
In the meantime, it is useful to ask what 
combination of $(l_1,l_2,l_3)$ is most of the signal coming
from, i.e. what triangle configurations in the bispectrum dominate.  
In Fig.~\ref{fig:lensconfiguration}, 
as an example, we show the general behavior of individual
mode contribution to our signal-to-noise statistic $\chi^2$
as a function of $l_1$ and $l_2$.  Here, we have taken $l_3=300$,
which is representative of the their behavior in the $l_3$ range of interest.
Since $B^2(l_1,l_2,l_3)$ is zero when $l_1+l_2+l_3$ is odd we have omitted these values for plotting purposes.

The configuration dependences results from 
a combination of the secondary-lensing cross-correlation power
spectra and the CMB power spectrum. 
They are modulated by the
behavior of the Wigner-3$j$ symbol 
and variations in the noise term from the CMB power spectra 
As shown in Fig~\ref{fig:wig3j}, the
Wigner-3$j$ symbol also peaks at the lowest $l_1$ 
that satisfies the triangle rule 
$l_1= l_3-l_2$.  The CMB power spectrum noise introduces features
associated with the acoustic peaks at $l \ga 200$ 
(see Fig.~\ref{fig:dopplermp} and \ref{fig:dopplerszisw}). 

The SZ-lensing bispectrum gets  most of its contribution when $l_1
\sim l_3 - l_2$ with $l_1$ taking values in the intermediate $l$ range of
few hundreds where the cross-power peaks.  
This behavior also suggest why the SZ$^{\rm (nl)}$-lensing
bispectrum is not strongly enhanced even at high
$l_3$ values by nonlinear effects, as most of the contributions come
from intermediate $b_l^{\rm SZ\; (nl)}$ values, where nonlinear
enhancement is negligible.

For ISW-lensing, the cross-correlation term peaks towards
low $l$ values, and thus, its configuration dependence is such that
the contributions to bispectrum peaks when $l_1 \sim l_3 - l_2$ with $l_1$
in the range of few tens. This behavior continues out to high
$l_3$ values.  

The Doppler-lensing configuration
also shows a similar behavior but with contributions from somewhat smaller
$l_1 \la l_3 - l_2$ with $l_1$ again taking values 
close to the peak in the cross-correlation power spectrum. 
These subtle differences in configuration may assist in isolating and identifying
the various contributions to the bispectrum once it is observed.

\section{Ostriker-Vishniac Coupling: Derivation}
\label{sec:ostrikervishniac}
We consider the generation of CMB bispectra through coupling between
the second order reionized Ostriker-Vishniac (OV) effect and two other linear
sources of anisotropies.  Since the OV effect is itself a secondary anisotropy
source, we shall see that it only couples to other secondary anisotropy sources.
First, we present a general derivation of the coupling
between OV effect and secondary effects and in \S~\ref{sec:ovderivation}
and then turn to specific secondary effects such as the ISW, SZ and 
reionized Doppler effect.  

\subsection{General Considerations}
\label{sec:ovderivation}
The OV effect arises from the modulation of the Doppler effect by
density fluctuations which affect the probability of scattering
(\cite{OstVis86a} 1986a; \cite{Vis87} 1987). The OV temperature
fluctuations can be written as
\begin{eqnarray}
T^{\rm OV}(\hat{\bf n})&=&  \int d\rad
	g(r) {\bf n} \cdot {\bf v}(r,\bn r) \delta(r, \bn r) \nonumber\\
&=&-i \int d\rad g \dot{G} G
\int \frac{d^3{\bf k}}{(2\pi)^3} \int \frac{d^3{\bf k}'}{(2\pi)^{3}}
\nonumber \\
&&\times \delta({\bf k}-{\bf k}')\delta({\bf k'})
e^{i{\bf k}\cdot \hat{\bf n}\rad} \left[ \hat{\bf n} \cdot
{\bf D}({\bf k},{\bf k}') \right] \, ,
\end{eqnarray}
where,
\begin{equation}
{\bf D}({\bf k},{\bf k}') =  {1 \over 2}\left[ \frac{\veck - \veck'}{| \veck -
\veck'|^2} + \frac{\veck'}{k'^2} \right] \, ,
\end{equation}
is the projection vector that couples the density and velocity perturbations.

We can now expand out the temperature perturbation, $T^{\rm OV}$, into
spherical harmonics:
\begin{eqnarray}
a^{\rm OV \ast}_{lm} &=& i \int d\hat{\bf n}
\int d\rad\; (g\dot{G} G) 
\int \frac{d^3{\bf k_1}}{(2\pi)^3}\int \frac{d^3{\bf
k_2}}{(2\pi)^3}
\delta^\ast({\bf k_1})\delta^\ast({\bf k_2}) \nonumber \\
&&\times e^{-i({\bf k_1+k_2})
\cdot \hat{\bf n}\rad} \left[ \hat{\bf n} \cdot
{\bf D}({\bf k_1},{\bf k_2}) \right]
Y_l^{m}(\hat{\bf n}) \, ,
\end{eqnarray}
where we have symmetrizised by using $\veck_1$ and $\veck_2$
to represent $\veck$ and $\veck-\veck'$ respectively.
Now the dot product between  $\hat{\bf n}$ and the projection vector
is
\begin{equation}
\hat{\bf n} \cdot
{\bf D}({\bf k_1},{\bf k_2}) = {1 \over 2} \left( \frac{\hat{\bf n} \cdot
\veck_1}{k_1^2}+\frac{\hat{\bf n} \cdot \veck_2}{k_2^2} \right) \, ,
\end{equation}
with
\begin{equation}
\hat{\bf n} \cdot \veck = \sum_{m'} \frac{4\pi}{3} k
Y_1^{m'}(\hat{\bf n}) Y_1^{m'\ast}(\hat{\veck}) \, .
\end{equation}
Since the dot product is symmetric, we only consider one term here,
and multiply the result by 2. 
By using the Rayleigh expansion (Eq.~[\ref{eqn:Rayleigh}]), we can 
rewrite the multipole moments as 
\begin{eqnarray}
a^{\rm OV \ast}_{lm} &=& i \frac{(4 \pi)^3}{3}
\int d\rad
\int \frac{d^3{\bf k}_1}{(2\pi)^3} \int \frac{d^3{\bf
k}_2}{(2\pi)^3} 
\sum_{l_1 m_1}\sum_{l_2 m_2}\sum_{m'} \nonumber\\
&& \times
(-i)^{l_1+l_2}
(g\dot{G}G) 
\frac{j_{l_1}(k_1\rad)}{k_1}
j_{l_2}(k_2\rad) 
\delta^\ast({\bf k_1})\delta^\ast({\bf k_2}) 
\nonumber\\
&& \times
Y_{l_1}^{m_1\ast}(\hat{\veck}_1) Y_1^{m'\ast}(\hat{\veck}_1)
Y_{l_2}^{m_2\ast}(\hat{\veck}_2)  \nonumber \\
&& \times \int d\hat{\bf n}
Y_l^{m}(\hat{\bf n}) Y_{l_1}^{m_1}(\hat{\bf n})
Y_{l_2}^{m_2}(\hat{\bf n})
Y_1^{m'}(\hat{\bf n})\,. 
\end{eqnarray}
The power spectrum of the OV effect may be calculated from
this expression.  Since the end expression is cumbersome
we revert to the expressions of \cite{Hu99} (1999) in calculating the
OV power spectrum shown in Fig.~\ref{fig:cl}.

To leading order in $\Delta^2(k)$, bispectrum contributions
involve one OV term and two linear sources of anisotropies.
Recall that a general linear effect 
can be expressed as 
a weighted projection of the density field (see \S \ref{sec:generallensing})
\begin{eqnarray}
a^{\rm S}_{lm} &=& i^l \int \frac{d^3\veck}{2 \pi^2}
\delta(\veck)  I_l^{\rm S}(k) \Ylmn(\hat{\veck}) \, , \nonumber \\
I_l^{\rm S}(k)&=& \int d\rad  W^{\rm S}(k,\rad)j_{l}(k\rad) \, .
\end{eqnarray}

After some straightforward but tedious algebra, we can write
\begin{eqnarray}
a^{\rm S}_{l_1 m_1}a^{\rm S}_{l_2 m_2}a^{\rm OV\ast}_{l_3 m_3} &=&
(4\pi)^2 \int \frac{dk_1}{k_1} \int \frac{dk_2}{k_2} \Delta^2(k_1) \Delta^2(k_2) 
I_{l_1}^{\rm S}(k_1) 
\nonumber \\
&&\times 
I_{l_2}^{\rm S}(k_2)  
[
I_{l_1,l_2}^\ov(k_1,k_2) +
I_{l_1,l_2}^\ov(k_2,k_1)] 
\nonumber\\
&&\times  \int d\hat{\bf n} Y_{l_3}^{m_3}(\hat{\bf n}) Y_{l_2}^{m_2*}(\hat{\bf n})
	Y_{l_1}^{m_1*}(\hat{\bf n})
\,,
\label{eqn:ovtriplet}
\end{eqnarray}
where
\begin{eqnarray}
I^\ov_{l_1,l_2}(k_1,k_2) &=& \int d\rad W^\ov j_{l_2}(k_2\rad)
j'_{l_1}(k_1\rad) \nonumber\,,\\
W^\ov(k_1,r) &=& -{1 \over k_1} g\dot{G}G \,.
\end{eqnarray}
In simplifying the integrals involving spherical harmonics,
we have made use of the properties of Clebsch-Gordon coefficients, in
particular, those involving $l=1$.

In order to construct the bispectrum, note that
\begin{equation}
\langle a_{l_1 m_1}^{\rm S} a_{l_2 m_2}^{\rm S} a_{l_3 m_3}^{\rm OV} \rangle
=
(-1)^{l_3}\langle a_{l_1 m_1}^{\rm S} a_{l_2 m_2}^{\rm S} a_{l_3
-m_3}^{\rm OV*} \rangle \; .
\end{equation}
Under the assumption that ``S'' denotes the sum of all the sources so that
the two contributions are indistinguishable, the bispectrum becomes
\begin{eqnarray}
B_{l_1 l_2 l_3} &=& \sum_{m_1 m_2 m_3} \wjm \Big(
\left< a^{\rm S}_{l_1 m_1}a^{\rm S}_{l_2 m_2}a^{\rm OV}_{l_3 m_3}  \right>
+ 
\nonumber\\
&& \times
\left< a^{\rm S}_{l_2 m_2}a^{\rm S}_{l_3 m_3}a^{\rm OV}_{l_1 m_1}  \right>
+
\left< a^{\rm S}_{l_3 m_3}a^{\rm S}_{l_1 m_1}a^{\rm OV}_{l_2 m_2}  \right> 
\Big)\nonumber\\
&=& \sqrt{\frac{(2l_1 +1)(2 l_2+1)(2l_3+1)}{4 \pi}}
\left(
\begin{array}{ccc}
l_1 & l_2 & l_3 \\
0 & 0  &  0
\end{array}
\right) \nonumber \\
&&\times [ b^{\se-\se}_{l_1,l_2} + {\rm Perm.}] \, .
\label{eqn:ovbidefn}
\end{eqnarray}
where we have used Eq.~(\ref{eqn:harmonicsproduct}), Eq.~(\ref{eqn:ortho}),
and Eq.~(\ref{eqn:negation}). 
Here,
\begin{eqnarray}
 b^{\se-\se}_{l_1,l_2}
&=&(4\pi)^2 \int \frac{dk_1}{k_1} \int \frac{dk_2}{k_2} \Delta^2(k_1) \Delta^2(k_2) \nonumber \\
&&\times I^\ov_{l_1,l_2}(k_1,k_2) I_{l_1}^{\rm S}(k_1) I_{l_2}^{\rm
S}(k_2) \, .
\label{eqn:finalintegral}
\end{eqnarray}
Note that we have rewritten the $k_1 \rightarrow k_2$ term
in Eq.~(\ref{eqn:ovtriplet})
as an $l_1 \rightarrow l_2$ interchange so that in Eq.~(\ref{eqn:ovbidefn}) 
``Perm.'' means a sum over the
remaining 5 permutations of ($l_1$,$l_2$,$l_3$) as usual.
In the following sections, we evaluate this expression with
\begin{equation}
I_l^\se(k) = I_l^\isw(k) + I_l^\sz(k) + I_l^\dop(k) + I_l^\sw(k).
\label{eqn:Iexpansion}
\end{equation}
The last term is the Sachs-Wolfe effect which we explicitly consider to 
show that the OV effect mainly couples with secondary not primary anisotropies.
We drop the double scattering effect of \S \ref{sec:doublelensing}
as it is a small contribution for small optical depths.
Note that expanding $I^\se$ in Eq.~(\ref{eqn:finalintegral}) produces
many cross-terms which we will call hybrid effects.

In general, Eq.~(\ref{eqn:finalintegral}) involves
five integrations,
three over radial distances and two over wavenumbers.
As in the lensing bispectrum calculation, these integrals can be simplified
using the Limber approximation for sufficiently large $(l_1,l_2)$.  
Here, we employ a version based on
the completeness relation of spherical Bessel functions
\begin{equation}
\int dk k^2 F(k) j_l(kr) j_l(kr')  \approx {\pi \over 2} \da^{-2} \deld(r-r')
						F(k)\big|_{k={l\over d_A}}\,,
\end{equation}
where the assumption is that $F(k)$ is a slowly-varying function.
Applying this to the integral over $k_2$ yields
\begin{eqnarray}
b^{\se-\se}_{l_1,l_2} &\approx& (4 \pi)^2 {\pi \over 2} \int dr \int dr_1 
\int {dk_1 \over k_1} {1 \over k_2^{3}}
\Delta^2(k_1) \Delta^2(k_2) \nonumber\\ 
&& \times W^{\rm S}(k_1,r_1) {W^{\rm S}(k_2,r) W^{\rm OV}(k_1,r) \over \da^{2}}
\Big|_{k_2 = {l_2 \over \da}}
						  \nonumber\\
&& \times j_{l_1}'(k_1 r) j_{l_1}(k_1 r_1)\,,
\label{eqn:velocityintegral}
\end{eqnarray}
Integrating by parts and assuming negligible boundary terms 
yields,
\begin{eqnarray}
b^{\se-\se}_{l_1,l_2}  &\approx& - 4 \pi^4 \int 
		dr {1 \over k_1^{4} k_2^{3} }
		\Delta^2(k_1) \Delta^2(k_2)  
		{ W^\se(k_1,r) \over \da^2}
			\nonumber\\
	   && \times
		{d \over dr} 
		\left[{W^\ov(k_1,r) W^\se(k_2,r)  \over \da^2}
			     \right]
				\Big|_{
				k_1={l_1\over \da},
				k_2={l_2\over \da}} \nonumber\\
	   &\approx&  4 \pi^4 \int 
		dr {1 \over  k_1^4 k_2^{3}} 
		\Delta^2(k_1) \Delta^2(k_2)  
		{d \over dr} 
		  \left[ {W^\se(k_1,r) \over \da^2} 
			\right]\nonumber\\
	   && \times
		{  W^\ov(k_1,r) W^\se(k_2,r)  \over \da^2}
			     \Big|_{
				k_1={l_1\over \da},
				k_2={l_2\over \da}} \,.
\label{eqn:finallimberintegral}
\end{eqnarray}
This Limber approximation reduces the dimension of the integrals
from 5 to 1.  Note that where the Limber approximation
applies only equal time correlations contribute.  On small
angular scales then only secondary, not primary, 
anisotropies couple to the OV effect in the bispectrum.

\subsection{ISW-ISW-OV coupling}

Recall that from \S~\ref{sec:ISWlensing},
the weight function for the ISW effect is
\begin{equation}
W^{\rm ISW}(k,\rad) = -\frac{3\Omega_m H_0^2}{k^2} \dot{F} \, .
\end{equation}
Substituting this weight function into Eq.~(\ref{eqn:finallimberintegral}) leads to
\begin{eqnarray}
b^{\isw-\isw}_{l_1,l_2} &=&  
-\frac{36 \pi^4}{l_1^7l_2^5}
\Omega_m^2 H_0^4
\int_0^{\rad_0} d\rad 
\Delta^2\left(\frac{l_1}{\da}\right)
\Delta^2\left(\frac{l_2}{\da}\right) \nonumber \\
&&\times 
\da^{10} 
g\dot{G}G \left({d \over d\rad}{\dot{F} \over \da^2}\right)
\dot{F}\;. 
\end{eqnarray}
We employ this Limber approximation in our calculation of the
ISW-ISW-OV bispectrum effect.

\subsection{SZ-SZ-OV}
In a similar manner, we can calculate the SZ-SZ-OV effect
following the discussion in \S \ref{sec:SZlensing} 
where the weight function is given as
\begin{equation}
W^\sz(k,\rad) = -2A^\sz a g G
\, ,
\end{equation}
and is independent of the wave vector $\veck$. 
Now the Limber approximated bispectrum term, used in the calculations 
presented here, is
\begin{eqnarray}
b^{\sz-\sz}_{l_1,l_2}&=&-\frac{16 \pi^4}{l_1^5 l_2^3} (A^\sz)^2 
\int_0^{r_0} d\rad 
\Delta^2\left(\frac{l_1}{\da}\right) \Delta^2\left(\frac{l_2}{\da}\right)
\nonumber \\
&& \times 
a g^2 G^2 \dot{G} 
\da^6 \left[ \frac{d}{d\rad} \frac{a g G}{\da^2} \right] 
\, .
\end{eqnarray}
%
%

Similar to our calculation on the enhancement of SZ-lensing bispectrum
due to nonlinear growth of density fluctuations, we also consider the
effect of nonlinearities on the SZ-SZ-OV bispectrum. As shown in
Fig.~\ref{fig:cl}, non-linearities enhance both SZ and OV effects and
at high $l$, the OV effect involves the large-scale velocity field and small
scale density field in the nonlinear regime (\cite{Hu99} 1999).
For the bispectrum coupling, therefore, we replace only one
of the power spectrum terms with the nonlinear relation,  
\begin{eqnarray}
b^{{\rm SZ(nl)}-\sz}_{l_1,l_2}&= &-\frac{16 \pi^4}{l_1^5 l_2^3}
(A^\sz)^2
\int_0^{r_0} d\rad 
\Delta^2\left(\frac{l_1}{\rad}\right) 
\Delta^{2\, {\rm (nl)}}\left(\frac{l_2}{\rad},r\right) \nonumber \\
&& \times 
\da^6
\left[ \frac{d}{d\rad}\frac{a g G}{\da^2} \right]  a g^2 \dot{G} \, .
\end{eqnarray}

As before, with SZ-lensing nonlinear cross-correlation, the SZ-SZ$^{\rm
(nl)}$-OV coupling is calculated with the nonlinear density power
spectrum evaluated along the line of sight at the corresponding
lookback time.

\subsection{Doppler-Doppler-OV}

The reionized Doppler effect was discussed in detail in \S \ref{sec:dopplerlensing}
and has a weight function
\begin{equation}
W^\dop(k,r)= {1 \over k^2}(\dot g \dot G + g \ddot G) \,.
\end{equation}
Unlike ISW-ISW-OV and SZ-SZ-OV, where the Limber
approximation is sufficiently accurate for our purposes,
explicit integration over wavenumber is 
necessary here.  
The Limber approximation breaks
down in the large-angle regime and hence affects the coupling between
the velocity field in the OV effect and the Doppler effect.
On the other hand, the density field in the OV effect is dominated
by small-scale fluctuations where the Limber approximation is excellent.
This implies that we may use
Eq.~(\ref{eqn:velocityintegral})
in calculating the Doppler-Doppler-OV effect:
\begin{eqnarray}
b^{\dop-\dop}_{l_1,l_2}&= & -\frac{8 \pi^3}{l_2^5}
\int \frac{dk}{k^2} \Delta^2(k) \int d\rad 
\Delta^2\left(\frac{l_2}{\da}\right) \nonumber \\
&&\times g\dot{G}G \da^3 \left(\ddot{G} g +\dot{G}\dot{g}\right)
j'_{l_1}(k\rad) I_{l_1}^{\rm dop}(k) \,.
\end{eqnarray}
The integrals above include a derivative of the Bessel function and
is numerically difficult to evaluate. As in 
Eq.~(\ref{eqn:finallimberintegral}), one can integrate by parts to
obtain a more tractable form.  
We considered both these approaches and the resulting bispectra 
agree at a level of 10\%, with most of the
difference resulting from the numerical computation of the coupling
term involving the derivative of the Bessel function. 
This agreement also suggests that the evaluation at $k=l/\da$ in the
Limber approximation can be performed after all the integration
by parts are complete as is assumed in Eq.~(\ref{eqn:finallimberintegral}).

\begin{figure*}[t]
\centerline{\psfig{file=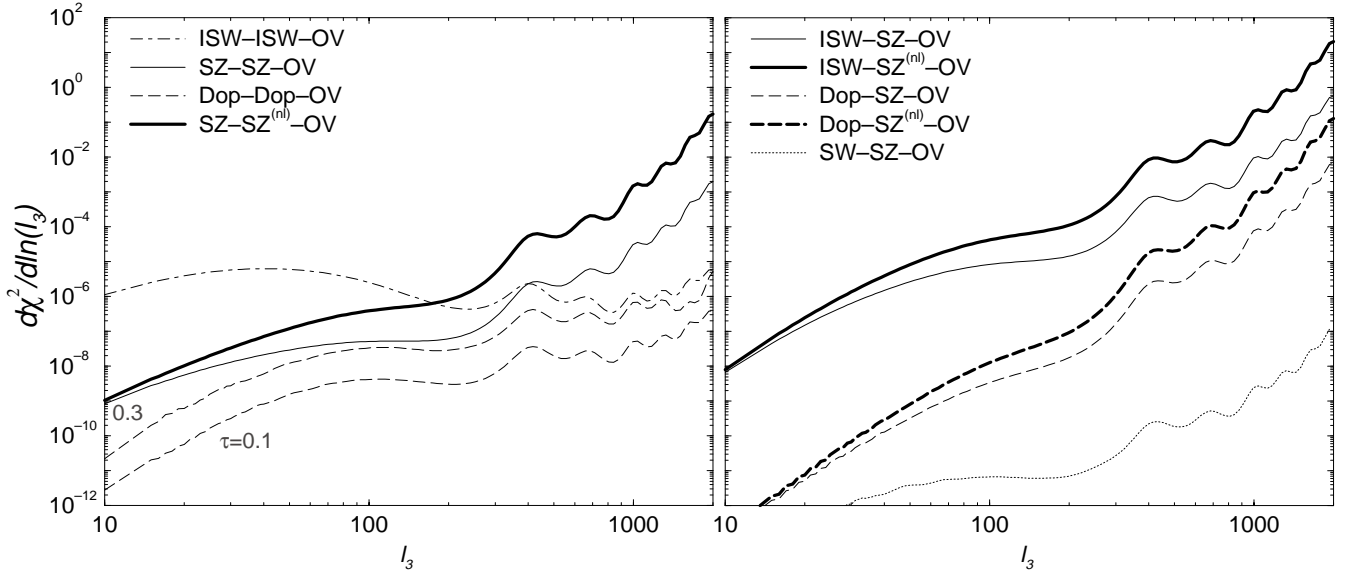,width=7.0in,angle=0}}
\caption{Contribution to $\chi^2$ per log interval in $l_3$ for the OV 
coupling effects for the fiducial $\Lambda$CDM model with
$\tau=0.1$, $\Delta z= 0.1(1+z_\ri)$ and no instrumental noise. 
{\it Left panel---} Coupling with two secondary effects
of the same kind. {\it Right panel---} Hybrid effects involving
secondaries of different kind.
Other than Doppler-Doppler-OV and Doppler-SZ-OV couplings, the
secondary-secondary-OV bispectra are only mildly sensitive to the
reionized optical depth and width.} 
\label{fig:chisqov}
\end{figure*}

\subsection{ISW-SZ-OV}

In addition to coupling between OV effect and secondary anisotropies of
similar kind, we also consider hybrid couplings.
The hybrids that contribute the most involve a large
scale effect such as the ISW effect to couple with the velocity field
in the OV effect, and a small-scale effect such as the SZ effect to couple with
the density field in the OV effect, 
\begin{eqnarray}
b^{\isw-\sz}_{l_1,l_2} &=&  
-\frac{24 \pi^4}
{l_1^7 l_2^3}
 \Omega_m H_0^2 A^\sz 
\int d\rad \Delta^2\left(\frac{l_1}{\rad}\right)
\Delta^2\left(\frac{l_2}{\rad}\right) 
\nonumber \\
&& \times
G^2 g^2 a \dot{G} \da^8 
\left(\frac{d}{d\rad}\frac{\dot{F}}{\da^2}\right) \,.
\end{eqnarray}
%
%
To estimate the effects of nonlinearities, we take the
approximations introduced for the SZ-SZ$^{\rm (nl)}$-OV 
above and find
\begin{eqnarray}
b^{\isw-\sz{\rm (nl)}}_{l_1,l_2} &=&  
-\frac{24 \pi^4}
{l_1^7 l_2^3}
 \Omega_m H_0^2 A^\sz 
\int d\rad \Delta^2\left(\frac{l_1}{\rad}\right)
\Delta^{2({\rm nl})}\left(\frac{l_2}{\rad},r\right) 
\nonumber \\
&& \times
g^2 a \dot{G} \da^8 
\left(\frac{d}{d\rad}\frac{\dot{F}}{\da^2}\right) \,.
\end{eqnarray}

\subsection{Doppler-SZ-OV}

Similar to ISW-SZ-OV effect, the hybrid coupling of 
the reionized Doppler and SZ effects to OV effect.
As with the Doppler-Doppler-OV calculation, we integrate the
coupling between Doppler effect and OV velocity part and use the
Limber approximation to describe the coupling between SZ and OV
density part. The bispectrum term is then,
\begin{eqnarray}
b^{\dop-\sz}_{l_1,l_2}&= & \frac{16 \pi^3}{l_2^3} A^\sz
\int \frac{dk}{k^2} \Delta^2(k) \int d\rad 
\Delta^2\left(\frac{l_2}{\da}\right) \nonumber \\
&&\times G^2 g^2 a \dot{G}\da j'_{l_1}(k\rad) I_{l_1}^{\dop}(k) \,.
\end{eqnarray}
As with the other OV-SZ couplings, the nonlinear generalization
of the effect is given by 
\begin{eqnarray}
b^{\dop-\sz{\rm (nl)}}_{l_1,l_2}&= & \frac{16 \pi^3}{l_2^3} A^\sz
\int \frac{dk}{k^2} \Delta^2(k) \int d\rad 
\Delta^{2(\rm nl)}\left(\frac{l_2}{\da},r\right) \nonumber \\
&&\times  g^2 a \dot{G}\da j'_{l_1}(k\rad) I_{l_1}^{\dop}(k) \,.
\end{eqnarray}

\subsection{SW-SZ-OV}

Finally, we consider a hybrid involving the Sachs-Wolfe 
(SW) effect at the last scattering surface and the SZ effect. 
The weight function for the SW
effect can be written as
\begin{equation}
W^{\rm SW}(k,\rad_*) = -\frac{\Omega_m H_0^2}{2k^2} F\delta^D(\rad-\rad_*)\, ,
\end{equation}
and is evaluated at the last scattering surface, $\rad_* \approx \rad_0$.
The Limber approximation implies that there is no coupling between the
SW effect and the SZ or OV effects because there is no overlap in their
weight functions.  To the extent, that such a coupling exists the Limber
approximation fails.

Since 
it is at large scales that the Limber approximation breaks down, 
we are interested in the coupling between OV velocity part 
and the SW effect. 
As in the Doppler effects, we 
use Eq.~(\ref{eqn:velocityintegral}) to evaluate the deviations from the
Limber approximation,
\begin{eqnarray}
b^{{\rm SW}-\sz}_{l_1,l_2} &=& 
-\frac{8 \pi^3}{l_2^3}
\Omega_m H_0^2  A^\sz
F(\rad_*) j_{l_2}(k \rad_*) 
\int \frac{dk}{k^4} \Delta^2(k)
\nonumber\\
&&\times 
\int_0^{r_0}   d\rad \Delta^2\left(\frac{l_2}{\rad}\right) \da
a g^2 \dot{G}G^2
j'_{l_1}(k\rad)\,. 
\end{eqnarray}
and its nonlinear analogue
\begin{eqnarray}
b^{{\rm SW}-\sz{\rm (nl)}}_{l_1,l_2} &=& 
-\frac{8 \pi^3}{l_2^3}
\Omega_m H_0^2  A^\sz
F(\rad_*) j_{l_2}(k \rad_*) 
\int \frac{dk}{k^4} \Delta^2(k)
\nonumber\\
&&\times 
\int_0^{r_0}   d\rad \Delta^{2{\rm(nl)}} 
\left(\frac{l_2}{\rad},r\right) \da
a g^2 \dot{G}
j'_{l_1}(k\rad)\,. 
\end{eqnarray}

Here, we do not consider couplings such as SW-SW-OV effect as the OV
density part, with a redshift window at low redshifts, does not couple
to SW effect at the last scattering surface.

\section{Ostriker-Vishniac Couplings: Results}
\label{sec:ovresults}

\begin{figure}[t]
\centerline{\psfig{file=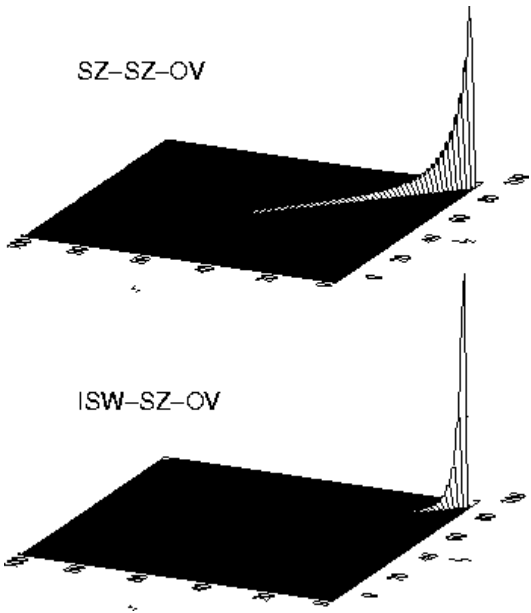,width=2.75in,angle=0}}
\caption{Configuration dependence of the SZ-SZ-OV ({\it top})
and ISW-SZ-OV ({\it bottom}) couplings with $l_3=100$.
This figure is analogous to the one shown in Fig.~\ref{fig:lensconfiguration} 
for secondary-lensing coupling. 
Contributions generally peak towards $l_3 \sim l_2$ and low $l_1$ for all $l_3$ values.} 
\label{fig:ovconfiguration}
\end{figure}

In Fig.~\ref{fig:chisqov}, 
we show $\chi^2$ contributions per logarithmic interval of $l_3$ 
for the coupling between
linear secondary effects and the OV effect as derived 
in the previous section.
Here, we have assumed cosmic variance in the noise only. 
For the coupling between two secondaries of the same type,
the bispectrum contributions per $l_3$ 
are substantially smaller than their 
lensing counterparts (see Fig.~\ref{fig:dopplerszisw}).
An examination of the configuration dependence of the contributions
shows that most of the contributions 
come from triplets which makes triangles with $l_1 \ll l_2 \lesssim
l_3$ (see Fig.~\ref{fig:ovconfiguration} for an example). 
The reason for this behavior is that the OV effect involves the large
scale bulk velocity field and small scale density fluctuations.  One 
secondary effect couples to the former and one to the latter creating
the desired configuration.  Because secondary effects tend to be strongly
peaked to either small or large angular scales, any coupling involving
two secondaries of the same type suffers suppression on one end or the
other.

Such configuration dependence also suggests that hybrid couplings should
contribute more strongly.  In particular in the last section, we considered
hybrids composed of one large-scale secondary effect coupling to the OV
velocity field and one small-scale secondary effect coupling to the OV
density field.
As shown in Fig.~\ref{fig:chisqov}, 
the contribution to the bispectrum produced by the hybrid ISW-SZ coupling
is substantially greater than either the ISW-ISW or SZ-SZ 
couplings. The same is true for the
Doppler couplings: Doppler-SZ coupling yields a larger effect than either
Doppler-Doppler or SZ-SZ coupling.  It is however a smaller contribution
than the ISW-SZ coupling.

Except for couplings involving the Doppler effect, the results
are fairly insensitive to the
reionized optical depth since the others are weighted toward low
redshifts. Increasing the optical depth to 0.3 from 0.1,
as shown in Fig.~7, the Doppler-Doppler-OV effect is increased by
factor of $\sim$ 10, and when the optical depth is further increased
to $\sim$ 0.5, the Doppler-Doppler-OV contribution becomes comparable to
that of the ISW-ISW-OV bispectrum signal at $l_3$ of a few thousand. 
The hybrid Doppler-SZ-OV bispectrum is less sensitive to the reionized optical depth than the
symmetric Doppler-Doppler-OV coupling, and is below the  bispectrum
produced by the hybrid ISW-SZ-OV for optical depths to reionization of
current interest. 

Nonlinearities in the density field can enhance couplings involving
the SZ effect.  Under the simplifying assumption that the gas density
traces the dark matter density, the enhancement in $\chi^2$ for
these effects is within a factor of 100, 
at $l_3$ of few thousands, which is consistent with the enhancement 
in the SZ and OV power spectra due to nonlinearities (see Fig.~\ref{fig:cl}), 
and the behavior of $l_1$, $l_2$ and $l_3$
configurations in producing  the OV bispectrum. The smaller 
enhancement factor in 
the SZ$^{\rm (nl)}$-lensing bispectrum, when compared to secondary-SZ$^{\rm
(nl)}$-OV effects, comes from the fact that the
secondary-lensing cross-correlation is not enhanced by non-linear
effects out to $l$ of $\sim$ 100 (see, Fig.~\ref{fig:dopplerszisw}), while
the cross-correlation between SZ and density part of the OV effect is
even enhanced at very low $l$ values, as can be seen from Fig.~\ref{fig:cl}.
One should note that the current calculation using the
non-linear power spectrum should be taken as an estimate of the
upper limit of the effects since on the smallest scales
gas pressure will make its distribution smoother than that of 
the dark matter (see discussion in 
\cite{Hu99} 1999).

In general, Limber approximation when applied to coupling between
OV effect and linear sources of an\-iso\-tro\-pies suggests that coupling
only exists for effects with strong temporal 
overlap in the weight functions $W^\se$ and $W^\ov$. 
This suggests that OV coupling with all primary effects, which contribute
at $z \sim 1000$ are heavily suppressed.
We tested this by numerically integrating the coupling to to the
SW effect.
As shown in Fig.~\ref{fig:chisqov},
the mismatch in redshift between the OV effect
and the SW leads to a large suppression of bispectrum signal and the
contribution is generally the lowest of all secondary-secondary-OV
couplings considered here.

\begin{figure}[t]
\centerline{\psfig{file=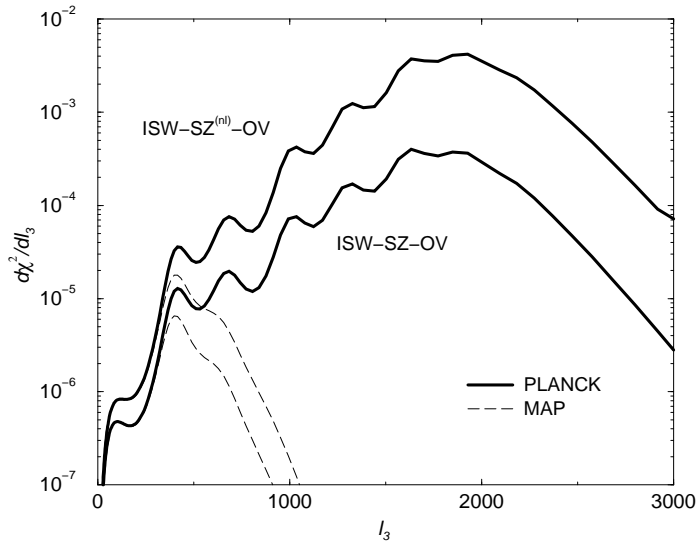,width=3.6in,angle=-90}}
\caption{Contributions to the $\chi^2$ as a function of $l_3$ for our fiducial
$\Lambda$CDM model for ISW-SZ-OV and its non-linear analogue
ISW-SZ$^{\rm (nl)}$-OV with MAP and
Planck noise included in the variance. As shown, the signal-to-noise ratios
are generally below the detection level for MAP and near or below the 
threshold for Planck ($\chi^2 \sim 0.3$,$3$ 
for the linear and non-linear effects).
These effects are the largest of the secondary-OV couplings considered.  }
\label{fig:ovmp}
\end{figure}

In Fig.~\ref{fig:ovmp},  we study the possible detection of 
OV-secondary couplings
by future satellite CMB missions. Here, we have only focussed on the
hybrid coupling of ISW and SZ effects to OV
effect (ISW-SZ-OV), and its non-linear analogue ISW-SZ$^{\rm
(nl)}$-OV, as they have the largest bispectrum signal of all secondary
anisotropies that couple to OV effect. The signal-to-noise ratio
generated by such effects is, however, small and their detection is
marginal at best even with Planck.  The ISW-SZ$^{\rm (nl)}$-OV
effect has the largest $\chi^2$ with a total value of $\sim$ 2.8 for
our fiducial $\Lambda$CDM cosmological model. In 
a flat universe with a cosmological constant, for reasonable variations
in $\Omega_m$ from 0.2 to 0.7, more or less consistent with current 
observational constraints on this parameter,  the total 
$\chi^2$ does not vary significantly to make this effect certainly
detectable with Planck. The variation in $\chi^2$ with $\Omega_m$ 
is such that the largest $\chi^2$ value is $\sim$ 3.1 when $\Omega_m
\sim 0.3$ and falls below 1 when $\Omega_m \gtrsim 0.7$ and tends to
zero with $\Omega_m \rightarrow 1$; the fall-off
is primarily due to the decrease in ISW effect with increasing
$\Omega_m$ and the ISW effect is zero in an Einstein-de Sitter
universe with $\Omega_m=1$.

Beyond MAP and Planck, 
the largest contribution to the signal-to-noise for a perfect experiment
comes from $l_3$'s corresponding to arcminute scales and
are at a level that eventual detection may be
possible. Even with such a high resolution experiment,
one should bear in mind that their detection requires configurations
in $l$ space with one short side which picks up contributions from 
the ISW effect.   
Its detection perhaps might involve a combination of a small angular scale
experiment such as the planned ground-based interferometers and 
the satellite missions which have the larger angular coverage.
The tight localization of 
the bispectrum contributions of modulated Doppler effects such as the OV effect
is also a useful property from the perspective of their role
as a contamination for the 
secondary-lensing measurements or intrinsic nongaussianity limits.
Even if the amplitude is enhanced, say by contributions from the
analogous patchy-reionization effect (see \S \ref{sec:power}), a large
part of the contamination may be removed by eliminating only a small
range in the bispectrum terms.   

\section{Discussion}
\label{sec:discussion}

Let us summarize the results of this study.
Gravitational lensing angular excursions couple with first order secondary
anisotropies generated at reionization to produce a bispectrum in the
CMB. Here, we have consider the coupling between lensing and the Doppler
effect and have shown this to be a significantly smaller effect on
the bispectrum than that between lensing and SZ and ISW effects 
for the low optical depths ($\tau \la 0.3$) expected in 
adiabatic CDM models. For the currently favored $\Lambda$CDM model, under the
assumptions made here on the physical state of baryonic gas distribution,
the bispectrum produced by  SZ-gravitational lensing angular
deflections is generally higher than that produced by ISW-lensing 
(\cite{GolSpe99} 1999).
Results from higher resolution hydrodynamic simulations and constraints from
observations are clearly needed to improve the calculation associated
with SZ effect.

The OV effect couples with two secondary
sources of anisotropies to produce a bispectrum. Given that OV effect
is due to a product of density and velocity fields, the
bispectrum contributions are maximized when one of the secondaries
peaks at large angular scales and the other at small angular 
scales. Such hybrid couplings produce relatively large
signals in particular the ISW-SZ-OV effect if the assumptions involved in
estimating the SZ contribution prove to be correct. 
For reasonable models of reionization and currently favored
cosmological models, 
however, the
contribution to bispectrum by secondary-secondary-OV coupling is below
the level that can be expected to be detected by MAP and is marginal
at best for Planck.   Although these signals are unlikely then to be detected in
Planck, they do serve as a source of residual systematic errors for other
measurements of the bispectrum that can grow to be comparable to the cosmic
variance term around the arcminute scale.  
Removal of these contributions is facilitated by the strong localization
of these effects at
$l_3 \sim l_2$, $l_1 \la 10$.  Removal of a broader range of $l_1$ 
is undesireable since many of the interesting bispectrum effects are also
maximized for $l_3 \sim l_2$. 

In general, bispectrum contributions from the coupling of secondary
anisotropies depends not only on the intrinsic amplitude of the secondary
effects but also their overlap in redshift.  In the small-scale Limber 
approximation, only equal time correlations contribute.  Many of the effects 
considered here are substantially smaller than one would naively guess due 
to mismatch in the epochs at which the effects contribute their signal.

We have considered contributions to the CMB bispectrum
from reionization to leading order in the density fluctuations.  
This is appropriate for the angular scales probed by the upcoming satellite
missions.  On arcminute scales and below, CMB anisotropies will be
dominated by contributions from truly non-linear structures 
in the gas density, temperature and ionization state.   The nongaussianity
induced on these quantities by structure formation will be a rich field
for future studies as experiments begin to probe the subarcminute regime
of CMB anisotropies.

\acknowledgments
We acknowledge useful discussions with David Goldberg, 
Lam Hui, Lloyd Knox, Roman Scoccimarro,
David Spergel and
Matias Zaldarriaga. ARC is grateful to Michael Turner and John Carlstrom for helpful advice and financial
support. WH is supported by the Keck Foundation, a Sloan Fellowship,
and NSF-9513835. We acknowledge the use of CMBFAST (\cite{SelZal96} 1996).

\appendix

\section{Useful Properties of the Wigner-3$j$ Symbol}

Here, we review the properties of the Wigner-3$j$ symbol that
are useful for the derivations in the text.  
It is defined by its relation to the Clebsch-Gordan coefficient,
\begin{eqnarray}
\wjm = (-1)^{l_1-l_2-m_3}
	{\langle l_1 m_2, l_2 m_2 | l_3 {-m_3}\rangle
	\over (2 l_3+1)^{1/2}}\,. \nonumber \\
\end{eqnarray}
and as a consequence obeys the orthonormality 
relation
\begin{eqnarray}
\sum_{m_1 m_2} \wjm \wjmp{1}{2}{4} = {
\deld_{\ell_3 \ell_4} \deld_{m_3    m_4    } \over 2\ell_3+1}\,,
\nonumber \\
\label{eqn:ortho}
\end{eqnarray}
the even permutation relation 
\begin{eqnarray}
\wjm = \wjmp{2}{3}{1} = \wjmp{3}{2}{1}\,.
\end{eqnarray}
the odd permutation relation
\begin{eqnarray}
\wjm = (-1)^{L}\wjmp{2}{1}{3}\,,
\end{eqnarray}
where $L=l_1+l_2+l_3$. Given symmetry under spatial inversions
only bispectrum terms with even $L$ are non-zero.
Since parity forces $L$ to be even, all permutations 
and negation of the $m$'s are equal.
Furthermore, the angular momentum selection rules require 
$l_i \le |l_j-l_k|$ for all permutations
of the indices, and $m_1+m_2+m_3=0$.   

\begin{figure}[t]
\centerline{\psfig{file=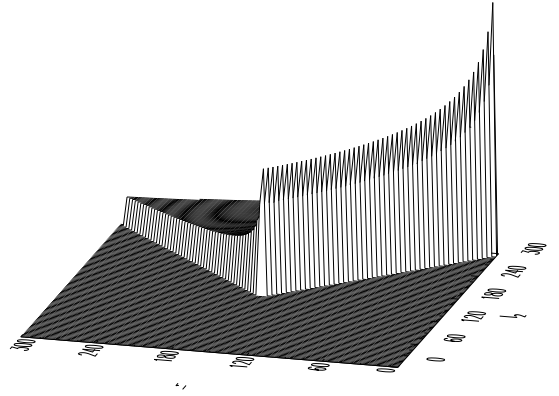,width=3.5in,angle=90}}
\caption{Absolute value of the Wigner-3$j$ symbol for $m_i=0$ as 
a function of $l_1$ and $l_2$ when $l_3=300$. 
We plot only even $l_1 +l_2 +l_3$.  }
\label{fig:wig3j}
\end{figure}

From the Clebsch-Gordan relation and series
one can derive the integral relation to 
spherical harmonics,
\begin{eqnarray}
\int d\bn
	\Ylm{1}
	\Ylm{2}
	\Ylm{3}
&=&
\sqrt{
	(2 l_1+1) (2 l_2+1) (2 l_3+1)\over 4\pi }\nonumber\\
&&\times
	\wj \wjm \,. \nonumber \\
\label{eqn:harmonicsproduct}
\end{eqnarray}
From conjugation of this relation, one finds that
\begin{eqnarray}
\wjm = (-1)^{L} \wjma{l_1}{l_2}{l_3}{-m_1}{-m_2}{-m_3}\,.
\label{eqn:negation}
\end{eqnarray}
For $m_i=0$, the Wigner-3$j$ symbol can 
be efficiently evaluated
\begin{eqnarray}
\wj &=&
(-1)^{L/2} \frac{({L \over 2})!}{({L \over 2}-l_1)!({L \over 2}-l_2)!({L \over 2}-l_3)!}
\nonumber \\
&\times&
\left[\frac{(L-2l_1)!(L-2l_2)!(L-2l_3)!}{(L+1)!}\right]^{1/2}
\nonumber \\
\end{eqnarray}
for even $L$; it vanishes for odd $L$.  In Fig.~\ref{fig:wig3j}, we
show the absolute value of the Wigner 3-$j$ symbol when $m_i=0$ as
given above. This expression is encountered in all our calculations on
the bispectra produced by secondary effects.
Note that other Wigner $3j$ terms, especially when $m_i$ is not equal
to zero may be efficiently
evaluated through recursion relations and the WKB approximation at high
multipole argument (\cite{SchGor75} 1975).  

\begin{thebibliography}{99}
\frenchspacing

\bibitem[Aghanim et al.]{Aghetal96}
	Aghanim, N., Desert, F.X., Puget, J.L, \& Gispert, R. 1996, A\&A, 311 1

\bibitem[Banday et al.]{Banetal99}
	Banday, A.J., Zaroubi, S. \& Gorski, K.M., ApJ submitted
        (astro-ph/9908070)

\bibitem[Bromley \& Tegmark]{BroTeg99}
	Bromley, B. \& Tegmark, M. ApJL in press (astro-ph/9904254)

\bibitem[Bunn \& White]{BunWhi97}
       Bunn, E. F. \& White, M. 1997, ApJ, 480, 6

\bibitem[Carlstrom et al.]{Car96}
	Carlstrom, J. E., Joy, M., Grego, L. 1996, ApJ, 456, L75

\bibitem[Cen \& Ostriker]{CenOst99}
	Cen, R., \& Ostriker, J. P. 1999, ApJ, 514, 1

\bibitem[da Silva et al.]{daS99}
	da Silva, A. C., Barbosa, D., Liddle, A. R., Thomas,
P. A. 1999, MNRAS, submitted, astro-ph/9907224

\bibitem[Eisenstein \& Hu]{EisHu99}
        Eisenstein, D.J. \& Hu, W. 1999, ApJ, 511, 5


\bibitem[Falk et al.]{Falk93}
      Falk, T., Rangarajan, R., Frednicki, M. 1993, ApJ, 403, L1


\bibitem[Ferreira et al.]{Feretal98}
 	Ferreira, P.G., Magueijo, J. \& Gorksi, K.M. 1998, ApJ, 503, 1

\bibitem[Gangui \& Martin]{GanMar99}
	Gangui, A. \& Martin, J., MNRAS submitted
		(astro-ph/9908009)

\bibitem[Gangui et al]{Ganetal94}
        Gangui, A., Lucchin, F., Matarrese, S. \& Mollerach, S.
	1994, ApJ, 430, 447

\bibitem[Goldberg \& Spergel]{GolSpe99}
	Goldberg, D. M. \& Spergel, D. N. 1999, PRD, 59, 103002

\bibitem[Griffiths et al.]{Grietal99}
	Griffiths, L. M., Barbosa, D., Liddle, A. R. 1999, MNRAS, 308, 845.

\bibitem[Gruzinov \& Hu]{GruHu98}
	Gruzinov, A. \& Hu, W. 1998, ApJ, 508. 435 

\bibitem[Haiman \& Knox]{HaiKno99}
	Haiman, Z., \& Knox, L. 1999, in {\it Microwave Foregrounds,}
ed. A. de Oliveira-Costa \& M. Tegmark (ASP: San Fransisco), astro-ph/9902311

\bibitem[Hinshaw et al]{Hinetal95}
	Hinshaw, G., Banday, A.J., Bennett, C.L., Gorski, K.M.,
	\& Kogut, A 1995, ApJ, 446, 67

\bibitem[Heavens]{Hea98}
	Heavens, A. 1998, MNRAS, 299, 805

\bibitem[Hu]{Hu99}
	Hu, W. 1999, ApJ in press (astro-ph/9907103).

\bibitem[Hu \& White]{HuWhi96}
	Hu, W. \& White M. 1996, A\&A, 315, 33.

\bibitem[Kaiser]{Kai84}
	Kaiser, N. 1984, ApJ, 282, 374

\bibitem[Kaiser]{Kai92}
	Kaiser, N. 1992, ApJ, 388, 286


\bibitem[Kendall \& Stuart]{KenStua69}
	Kendall, M. G., \& Stuart, A. 1969, Advanced Theory of
Statistics, Vol. II (London: Griffin)


\bibitem[Knox]{Kno95}
  Knox, L. 1995, PRD, 48, 3502

\bibitem[Knox et al.]{KnoScoDod98}
  Knox, L., Scoccimarro, R., \& Dodelson, S. 1998, Phys. Rev. Lett.,
        81, 2004

\bibitem[Limber]{Lim54}
        Limber, D. 1954, ApJ, 119, 655

\bibitem[Luo]{Luo94}
	Luo, X. 1994, ApJ, 1994, ApJ, 427, 71

\bibitem[Luo \& Schramm]{LuoSch93}
	Luo, X. \& Schramm, D.N 1993, PRL, 71,1124

\bibitem[Ostriker \& Vishniac]{OstVis86a}
        Ostriker, J.P., \& Vishniac, E.T. 1986a, Nature, 322, 804

\bibitem[Pando et al.]{Pando98}
	Pando, J. Vallas-Gabaud D. \& Fang, L. 1998, PRL, 79, 1611

\bibitem[Peacock \& Dodds]{PeaDod96}
        Peacock, J.A. \& Dodds, S.J. 1996, MNRAS, 280, L19

\bibitem[Peebles]{Pee80}
        Peebles, P.J.E. 1980, The Large-Scale Structure of the
Universe,        (Princeton: Princeton Univ. Press)

\bibitem[Rees \& Sciama]{ReeSci68} Rees, M. J. \& Sciama, D. N. 1968,
Nature. 519. 611


\bibitem[Sachs \& Wolfe]{SacWol67} Sachs, R. K., \& Wolfe, A. M.,
1967, ApJ, 147, 73

\bibitem[Schulten \& Gordon]{SchGor75}
	Schulten, K. \& Gordon, R.G. 1975, J. Math. Phys., 16, 1971

\bibitem[Seljak]{Sel96}
	Seljak, U. 1996, ApJ, 460, 549

\bibitem[Seljak \& Zaldarriaga]{SelZal96}
	Seljak, U. \& Zaldarriaga, M. 1996,
	ApJ, 469, 437

\bibitem[Seljak \& Zaldarriaga]{SelZal99}
	Seljak, U. \& Zaldarriaga, M. 1999,
	preprint, astro-ph/9811123


\bibitem[Spergel \& Goldberg]{SpeGol99}
	Spergel, D. N. \& Goldberg, D. M. 1999, PRD, 59, 103001

\bibitem[Sunyaev \& Zel'dovich]{SunZel80}
        Sunyaev, R.A. \& Zel'dovich, Ya. B. 1980, MNRAS, 190, 413

\bibitem[Tegmark et al.]{Tegetal99}
	Tegmark, M., Eisenstein, D.J., Hu, W., de Oliveira-Costa, A.
	ApJ, in press

\bibitem[Vishniac]{Vis87}
        Vishniac, E.T. 1987, ApJ, 322, 597

\bibitem[Wang \& Kamionkowski]{WanKam99}
	Wang, L. \& Kamionkowski, M. 1999, preprint, astro-ph/9907431

\bibitem[White et al.]{Whietal99}
	White, M., Carlstrom, J.E., Dragovan, M. \& Holzapfel, W.L. 1999
	ApJ, 514, 12

\bibitem[Zaldarriaga \& Seljak]{ZalSel97}
        Zaldarriaga, M. \& Seljak, U. 1997, Phys. Rev. D. 55, 1830


\end{thebibliography}
\end{document}